\documentclass[3p,review,authoryear]{elsarticle}

\usepackage[T1]{fontenc}
\usepackage[utf8]{inputenc}
\usepackage{mathrsfs}
\usepackage{array}
\usepackage{amssymb, amsmath, amsthm}
\usepackage{graphicx}
\usepackage{lmodern,url}
\usepackage{makecell} 
\usepackage{multirow}
\usepackage{microtype}
\usepackage{lineno}
\usepackage{xspace}
\usepackage{xcolor}
\usepackage{siunitx}
\usepackage{todonotes}
\usepackage{longtable}
\usepackage{booktabs}
\usepackage{multirow}

\usepackage{txfonts}

\usepackage{varioref}

\usepackage{mathdots}
\usepackage{subcaption}
\newcommand{\ie}{\emph{i.e.}\xspace}
\newcommand{\ca}{\emph{ca.}\xspace}
\newcommand{\eg}{\emph{e.g.}\xspace}

\newcommand{\dd}{\mathrm{d}}

\newcommand{\ho}{h_0}
\newcommand{\muf}{\mu_f}
\newcommand{\rhof}{\rho_f} 		  
\newcommand{\rhos}{\rho_s} 		  

\newcommand{\TRE}{\ensuremath{\text{TRE}}\xspace}
\newcommand{\turb}{T}
\newcommand{\OF}{\text{OF}}
\newcommand{\EG}{\eta_g}
\newcommand{\qmax}{Q_{\text{max}}}
\newcommand{\qPNK}{Q_{\text{PNK}}}

\newcommand{\Gr}{\mathit{Gr}}
\newcommand{\NRe}{\mathit{Re}_s}
\newcommand{\NRed}{\mathit{Re}_H}
\newcommand{\ReCell}{R_x}
\newcommand{\ReCellb}{\tilde{R}}
\newcommand{\ba}{\tilde{b}}

\newcommand{\uscale}{\mathscr{U}}
\newcommand{\vscale}{\mathscr{V}}
\newcommand{\longscale}{\mathscr{L}}
\newcommand{\blscale}{\delta}

\newcommand{\oD}[2]{\frac{\mathrm{d}#1}{\mathrm{d}#2}}

\newcommand{\ioD}[2]{\mathrm{d}#1/\mathrm{d}#2}

\newcommand{\WI}{\vset}
\newcommand{\wSt}{\ensuremath{w_{\text{St}}}}
\newcommand{\vmean}{\bar{u}}

\newcommand{\ghat}{\boldsymbol{\hat{g}}}
\newcommand{\xhat}{\boldsymbol{\hat{x}}}
\newcommand{\yhat}{\boldsymbol{\hat{y}}}
\newcommand{\nhat}{\boldsymbol{\hat{n}}}
\newcommand{\bv}[1]{\mathbf{#1}}

\newcommand{\tp}{t_{\perp}}

\newcommand{\vset}{w}

\newcommand{\wparam}{\Lambda}
\newcommand{\SY}{S} 
\newcommand{\dimClar}{\hat{S}}
\newcommand{\rGR}{\Gamma}

\newcommand{\grad}{\boldsymbol{\nabla}}

\newcommand{\Cv}{\phi} 

\definecolor{cfi}{rgb}{0,0.2,0.6}
\definecolor{beaublue}{rgb}{0.74, 0.83, 0.9}
\definecolor{BB}{rgb}{0, 0, 0}

\newcommand{\tilt}{\theta} 

\newcommand{\Ri}[1]{\textcolor{cfi}{#1}} 
\newcommand{\RB}[1]{\textcolor{BB}{#1}}
\newcommand{\BES}{SIS\xspace}
\newcommand{\Vp}{\ensuremath{V_p}} 


\graphicspath{{Figures/}}

\begin{document}
%

\begin{frontmatter}
\title{A review on steeply inclined settlers for water clarification}

\author[AMTC]{Cristian Reyes}
\author[AMTC,UA]{Fernando Apaz}
\author[AMTC,DIC]{Yarko Niño}
\author[AMTC]{\RB{Belén Barraza}}
\author[Nordita]{Cristobal Arratia}
\author[AMTC,DIMIN]{Christian F. Ihle\corref{cor}}
\cortext[cor]{Corresponding author}
\ead{cihle@ing.uchile.cl}
\address[AMTC]{Advanced Mining Technology Center, University of Chile. Tupper 2007, Santiago, Chile}
\address[UA]{\RB{Department of Mechanical Engineering, Universidad de Antofagasta, Av. Angamos 601, 1240000 Antofagasta, Chile}}
\address[DIC]{Department of Civil Engineering and Advanced Mining Technology Center, University of Chile, Av. Blanco Encalada 2002, Santiago, Chile}
\address[Nordita]{Nordita, KTH Royal Institute of Technology and Stockholm University, SE-106 91 Stockholm, Sweden}
\address[DIMIN]{Laboratory for Rheology and Fluid Dynamics, Department of Mining Engineering, University of Chile. Beauchef 850, 8370448 Santiago, Chile}
\date{\today}

\begin{abstract} 
Steeply inclined settlers (\BES) are solid-liquid separators that feature close, inclined confining elements that allow significant settling enhancement when compared to vertical tanks, resulting in relatively low footprints at equal throughput. The present article reviews the working principle, flow configuration, capacity, and several technological advances related to this kind of equipment. The consistency of the Ponder, Nakamura and Kuroda theory, developed after the discovery of the Boycott effect in~1920, which is essential to the settling enhancement effect, and the engineering approach developed independently for the design of \BES settlers during the late 1960's, based on particle trajectory analysis, is established. A discussion about potential developments for future improvement of the technology is made, with emphasis on three main topical areas: improvements of settling element array, optimization of inlet conditions, and potential improvements of the design of settling elements. The application of the technology to the mining industry is discussed in the context of increasing water scarcity and the progressive ore grade decrease.
\end{abstract}

\end{frontmatter}


%

\section{Introduction}\label{s:intro}

Water footprint has become a central issue in mineral processing plants and is inherently related to the sustainability of the mining industry. The progressive intensification of water recycling processes and the projected spreading of mineral concentrator plants using saltwater~\citep{Ihle18JoCP} in a context of progressively decreasing ore grades~\citep{Northey14RCaR} imply a challenge not just to mineral recovery~\citep{Rao89ME}, but also to water recycling from tailing stream operations as well as related processes such as solid handling infrastructure because of factors including complex rheology issues~\citep{Merrill17ME,Contreras20ME} and long distances connecting sources and destinations for such increasingly fine and concentrated slurries~\citep{Ihle14JoCP}. Tailing systems, and in particular, their storage facilities, are presently the center of attention of the mining industry due to their implications on environmental impact~\citep{Lottermoser10book} and even the safety of the increasing number of people living in their surroundings~\citep{Franks21SR}. Their mineral characteristics exhibit strong variability, both between mining operations~\citep{Bulatovic99proc} and within them~\citep{Tungpalan15ME}, which is inherited to slurry and suspension properties, including settling velocity of the fine contents and the rheological properties of both the slurries and the sludge at the underflow of settling unit processes. Water recovery in mineral processing is mostly made in thickeners, which in most cases work indeed as thickener-clarifier units~\citep{Concha14book,Arjmand19JoWPE}. Thickening and clarification operational optima are not necessarily the same, and control measurements focused solely on setting a specific underflow concentration may not meet those to ensure minimum turbidity of the overflow~\citep{Castillo19M}. \citet{Castillo19M} confirmed the central role of turbulence intensity on the quality of the overflow, even without the addition of reagents other than those use in the previous (thickening) stage. The potential to detach thickening and clarification thus offers an opportunity to fine-tune both processes separately and may report benefits including process stability and reduced overall flocculant consumption. 

Steeply inclined settlers (named hereafter as \BES) are low-footprint, gravitational solid-liquid separators suitable for clarification purposes used either as stand-alone unit processes or in combination with others depending on the nature of the suspensions to be treated. They feature closely spaced sloping sections (either plates or conduits), with typical angles, measured from the horizontal, between~\SI{45}{\degree} and \SI{60}{\degree}~\citep{Smith13CES}, plate separations or conduit hydraulic diameters typically of a few centimeters (in some cases, below \SI{1}{cm}) and section lengths of about \SI{2}{m}~\citep{Wilson05book,Tarleton07book} thereby confining otherwise homogeneous solid-liquid feeds and enhancing the separation rate when compared to vertical ones such as clarifier tanks\footnote{Low inclination settlers, also referred to as \emph{essentially horizontal}, whose slight tilt (on the order of a few degrees from the horizontal) is primarily to facilitate backwash operations~\citep{Culp68JAWWA}, are excluded from the present definition of \BES.}. Belonging to the class of so-called high-rate settlers, \BES technology is most commonly used for water clarification purposes to remove turbidity from water, commonly in combination with coagulation and flocculation as pre-treatment operations, thus proving a cost-effective solution for secondary treatment in wastewater plants~\citep{Jimenez97ET,Clark09JoEE,Edzwald10book}. A~2005 survey reveals an installed clarification capacity using this working principle of over 4.1~million~\SI{}{m^3/day}, with extensive application in France~\citep{Wilson05book}. Such a survey does not include China, but \citet{Zhou12AMaM} point out the incorporation of several dozens of lamella thickeners. 

In the mining industry, the increasingly widespread appearance of clays in gangue minerals tends to hinder recycled water quality~\citep{Connelly11AJoM,Graefe17book}. \BES technology is a low footprint, natural option for secondary treatment of turbid water to decouple the thickening function from the overflow cleaning process objective in otherwise thickener-clarifier units. This additional process flexibility becomes increasingly convenient both from a process perspective and from an economic one, considering that increased dosing of high molecular weight flocculants~\citep{Connelly11AJoM}, or additional dilution prior flocculation when compared with tailings without significant content of such gangue minerals~\citep{Graefe17book}, might be required in the presence of clays. Although the use of \BES equipment is not new to the mining industry~\citep[see, \eg,][]{Fuerstenau03book,MO21web,Parkson21web,Westech21web}, a comparatively small number of large-scale mineral processing plants use this technology and mostly rely on thickeners to control overflow water turbidity. An additional advantage of splitting thickening and clarification refers to underflow handling: separating fine solids from the bulk of the thickener underflow also splits the concentrated slurry problem into one which is massive in throughput but with comparatively coarser material and a second one (coming from the clarifier underflow) which, in spite of being a fine slurry (and thereby potentially complex), would most likely have a significantly smaller throughput than that of the plant feed stream.
 

The present article focuses on \BES fundamentals and reviews advances on the knowledge of its working principle and related technology development. A discussion on the opportunities to increase the quality of overflow water in mineral processing plants using \BES equipment is made. The present perspective article complements relatively recent and more general reviews on gravity concentration and classification in mineral processing~\citep{Tripathy15PT,Das18MPaEMR,Galvin21CES}. 




\section{Overview}\label{s:be}

\subsection{\RB{Historical background}}
\begin{figure}[!h]
\centering
\includegraphics[width=16.2cm]{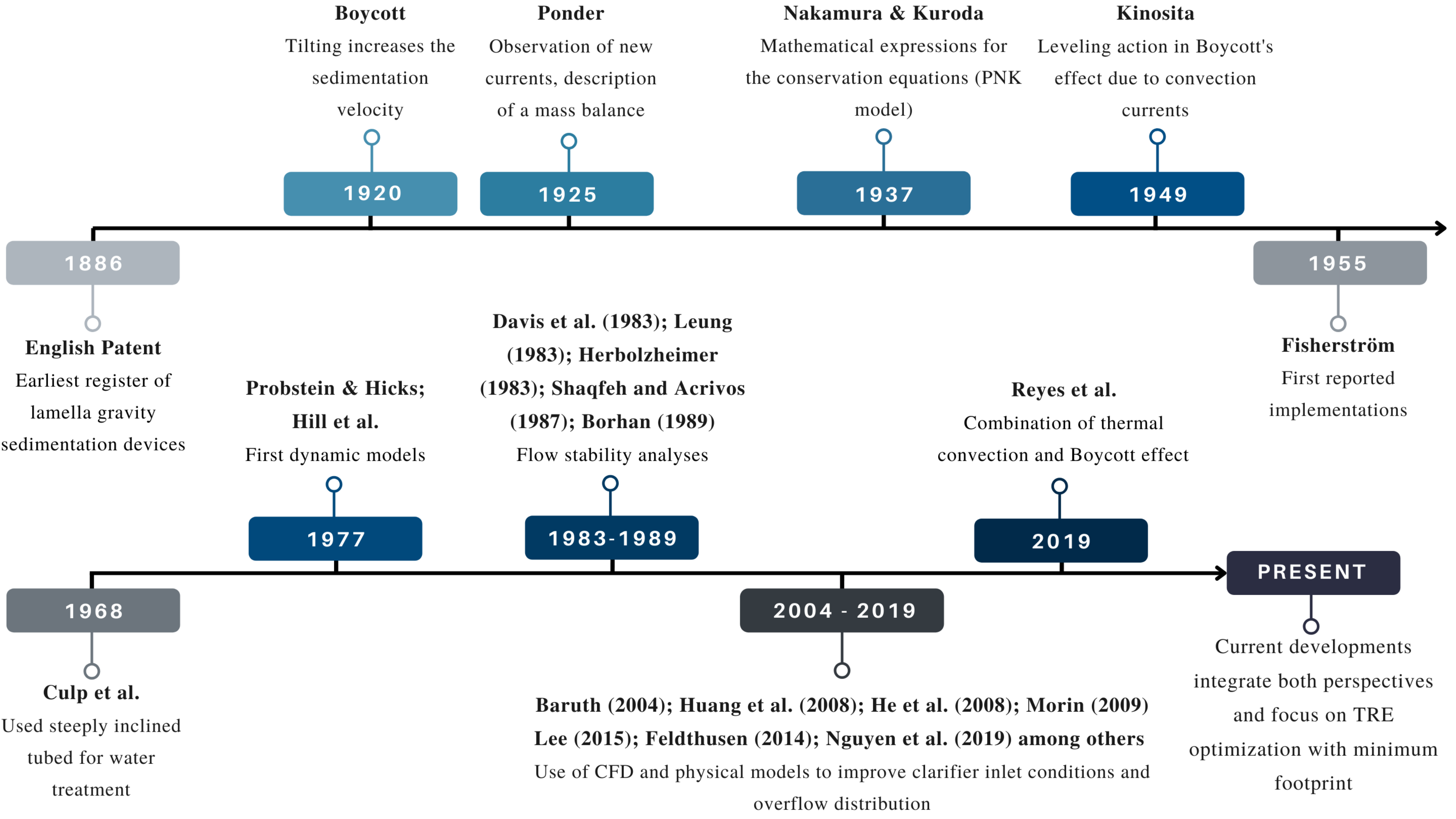}
\caption{\RB{Timeline showing the evolution of the SIS development from a scientific (top row, blue) and industrial (bottom row, gray) perspective.}}\label{f:timeline}
\end{figure}

\RB{The development of SIS has the peculiarity of being studied from two different outlooks in parallel: a theoretical-experimental perspective and an industrial perspective (see Figure \ref{f:timeline})}. The earliest record of the enhancement of settling efficiency by wall inclination was due to the British pathologist Arthur~Edwin~Boycott. 
Over 100~years ago, he discovered that the settling of red blood cells was enhanced by tilting test tubes from an otherwise upright position. 
In {a brief letter to {\it Nature}~\citep{Boycott20N}, he reported the proportions of clear serum after 5~hours of settling indicated in Table~\ref{t:Boycott}, where faster settling was found at a tilt angle $\tilt\approx\SI{56}{\degree}$ (measured from the horizontal). 
His tentative explanation of the phenomenon, based on an angle-dependent hindrance of sedimentation by Brownian motion, was not correct. 
Subsequent researchers would identify the release of clear fluid below the upper wall as the main mechanism in the increased sedimentation rate~\citep{Ponder25QJoEPTaI,Nakamura37KJM,Kinosita49JoCS}, which would later be convincingly confirmed by \citet[see also further references therein]{Hill77IJoMF}.  
Nevertheless, Boycott's observation} was the first step on what would be afterwards a whole research line both in fluid dynamics and process engineering, the so-called \textit{Boycott effect}, which is essential to \BES technology.

%
\begin{table}[!h]
\centering
\caption{Clear serum proportions (\%) reported by~\citet{Boycott20N} after 5~hours of settling.}\label{t:Boycott}
\begin{tabular}{ccccc}\toprule
Tube diam. & \multicolumn{4}{c}{Inclination from the horizontal}\\\cmidrule{2-5}
(mm) & \SI{90}{\degree} & \SI{78.75}{\degree} &\SI{67.5}{\degree} & \SI{56.25}{\degree}{}\\\midrule
2.7 & 6 & 20 & 29 & 51\\
8   & 5 & 10 & 15 & 21\\
14  & 4 & 5  & 9  & 12\\\bottomrule
\end{tabular}
\end{table}


In the water treatment industry, Boycott's discovery remained seemingly unnoticed for decades. 
While the earliest invention related to inclined plate separation appears to be an English patent from 1886 \citep{Forsell75JWPCF,Baruth04book}, the first reported implementations were made in Sweden in the 1950's (\citealp{fischerstrom1955sedimentation}; described by \citealp{hendricks2006water}), 
and the earliest engineering approaches to the design of tubular \BES date to about a decade later \citep{Culp68JAWWA}. 
With emphasis on reducing clarification equipment footprint~\citep{Edzwald10book}, these advances were inspired by earlier theories 
advocating the use of shallow settling basins to increase the capture surface \citep{Hansen67JAWWA,hendricks2006water}. 
This focus on increased sediment capture by the upward-facing plate can be contrasted with the earlier explanation of the Boycott effect, mentioned above, of increased release of clear fluid below the (downward-facing) upper wall; while not equivalent~\citep{Hill77IJoMF}, it will be shown that these two points of view are consistent and can under some conditions yield the same result. 
This difference in perspectives is a sign that, as it would remain the case for much subsequent work \citep[\eg,][and references therein]{De17book}, those first attempts were developed without any mention of Boycott's finding or subsequent research addressing the increase in clarified liquid production under inclined walls. 
In fact, the high sedimentation rate observed for high inclination angles came as a surprise along another key advantage of \BES. 
As summarized by \citet{Culp68JAWWA}, \textquotedblleft 
an angle of 60 deg provides continuous sludge removal while {\em still} allowing the tube to function as an efficient sedimentation device.\textquotedblright (emphasis added). The down-sliding of the concentrated settlement opens the way to sedimentation processes in continuous operation.

\subsection{\RB{Operation fundamentals}}
As advanced above, the operation of \BES can be classified according to whether the clarifier is open or closed (\ie, whether it operates in continuous or batch mode). Industrial operation of \BES equipment is mostly continuous~\citep[see, \eg,][for the case of the water treatment industry]{Edzwald10book}. 
In continuous operation, \BES are also classified mainly as countercurrent or cocurrent, but also possibly as cross-flow, according to the relative position between the feed and the underflow. 
While in countercurrent equipment the feed is located at or close to the bottom~\citep[the most common configuration in \BES equipment,][]{Wilson05book,Crittenden12book}, in cocurrent ones the feed is at or near the top~\citep[its use has been suggested for low concentration, network forming flocs by][]{Forsell75JWPCF}. 

\begin{figure}[!h]
\centering
\includegraphics[width=12cm]{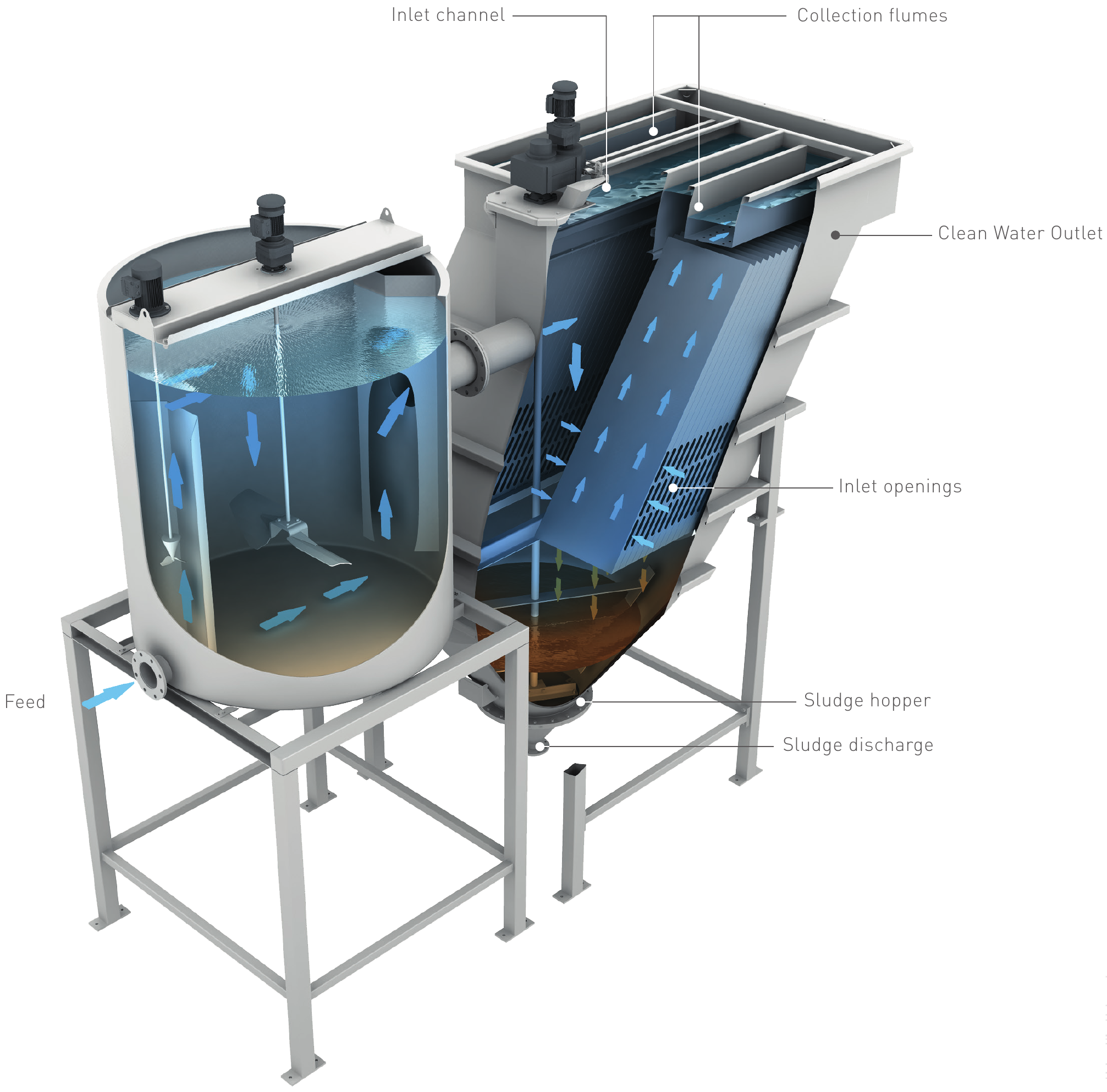}
\caption{Schematic of lamella, countercurrent settler with a flocculant tank upstream the separator (at the left of the figure). Image reproduced with permission of~\citet[][]{Nw21web}.}\label{f:lamella-countercurrent}
\end{figure}

\begin{figure}[!h]
    \centering
    \begin{tabular}{c}
        (a) \\ 
         \includegraphics[scale=1]{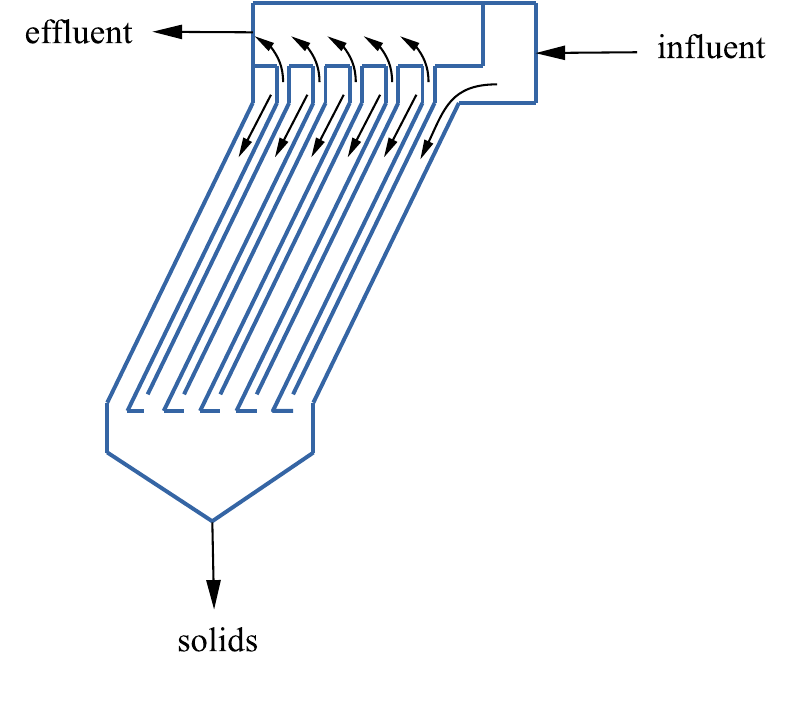}\\
        (b) \\
         \includegraphics[scale=1]{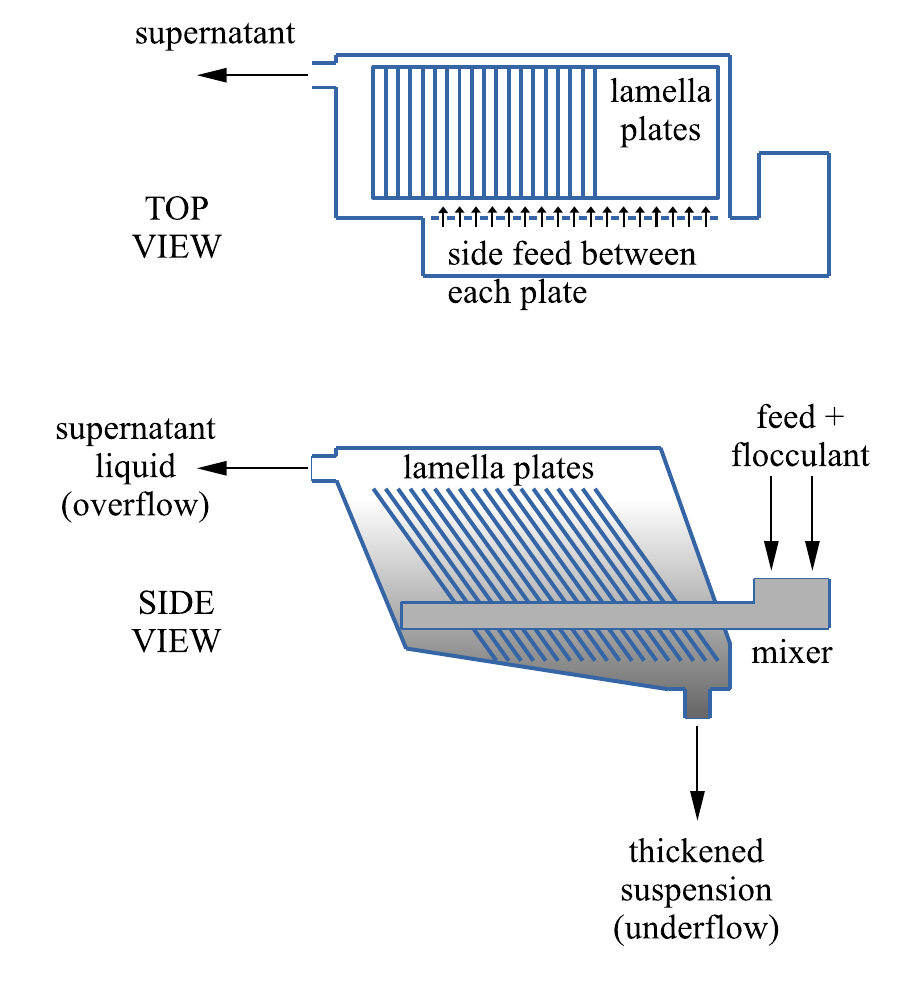}
    \end{tabular}
    \caption{Common configurations of \BES settlers. (a) Cocurrent flow~\citep[schematic adapted from][]{Crittenden12book}, and (b) cross-flow with side feed~\citep[schematic adapted from][]{Tarleton07book}.}
    \label{f:flow-config}
\end{figure}

Figure~\ref{f:lamella-countercurrent} shows a perspective view of a Lamella separator~\citep{Nw21web}, designed to operate in countercurrent configuration. 
It can be seen that the feed is located at the lowest portion of the array of inclined plates (the inlet openings on the sides), setting a clear fluid flow moving upwards (towards the overflow collection flumes) with a direction opposing to that of the sludge (moving downwards and collected in a hopper for subsequent discharge and transport). 
Figure~\ref{f:flow-config} shows diagrams of the other two possible configurations of this kind of equipment, namely cocurrent and cross-flow.  
In the cocurrent configuration (Figure~\ref{f:flow-config}a), the feed enters either at the top or at some intermediate point in the length of the settling element towards the bottom (underflow), without a net sideways component. 
In the cross-flow configuration (Figure~\ref{f:flow-config}b), there is an overall component of the flow that crosses settling elements sideways, where there is a countercurrent component due to the fact that the feed is below the overflow.

Inherent to the concept of \BES technology is the existence of inclined confining elements. They can be plate-like, thus defining lamella~\citep[also named inclined plate settlers or IPS,][]{Forsell75JWPCF} compatible with the cross-flow configuration, or have closed settling elements forming an array of conduits of square, rectangular, hexagonal, circular, or Chevron-like (V-shaped) cross-section conduits. Such packs of settling elements do not need to be set in a single direction, and there are cases where arrays of alternating direction sections are built, which can provide additional stability to the settler~\citep{Letterman99book}. The referred characteristic of an array of settling elements makes, at equal throughput, their footprint considerably smaller than that of vertical equipment. Examples of such differences in footprint and performance are numerous, including~\citet{Wenk90FR}, who reports, in the case of the phosphate industry, the use of \SI{10}{\%} of an otherwise vertical settler footprint at an equally effective settler area. In the context of low-cost technology utilization for turbidity removal of drinking water, \citet{Wisniewski13JoHE}  reported a footprint reduction up to \SI{76}{\%} compared to a vertical tank.  Also using plate-like elements, the numerical simulations of~\citet{Tarpagkou14PT} revealed an overall differential increase of \SI{18}{\%} in settling efficiency, adding a lamella settler bank to an existing primary settling tank. More recently, \citet{Wang20W} report a threefold increase in fine sediment removal efficiency of an array of inclined tubes compared with a vertical tank.


\section{\RB{Characteristics of steeply inclined settlers  (\BES)}}

\subsection{Surface overflow rate (SOR)}\label{s:SOR}

\BES technology excels in its low footprint when compared to other technologies for the same purpose. Key performance parameters of this technology are the surface overflow rate (SOR\footnote{To this purpose, the term surface loading is also used in the literature.}) and the effective surface area ($A_E$). They are defined, in terms of the geometrical parameters defined in Figure~\ref{f:PNK_schematic}:
\begin{equation}\label{e:SOR_definition}
    \text{SOR}= \frac{Q}{nA_E};\quad A_E=L W \cos\tilt,
\end{equation}
where $Q$ is the total overflow rate (in volume/time) $L$ and $W$ are the length and width of the plate, respectively, and $n$ is the number of vertical arrays of plates placed sideways. In other words, $nA_E$ is the total horizontally projected area of the settling system. Given a particular settler geometry and the characteristics of a suspension~\citep[including its temperature, see][for a discussion]{Takata17WSaTWS}, the particular value of the SOR (determined also by the inflow rate) will affect what part (or possibly~\SI{100}{\%}) of the suspended solids will be removed. \RB{In terms of engineering design guidelines, Table \ref{t:SOR} shows recommendations regarding SOR values for different types of sediments.}

%

%

\newcolumntype{P}[1]{>{\centering\arraybackslash}p{#1}}
\begin{table}[!h]
\centering
\caption{Design guidelines for SIS}
\label{t:SOR}
    \begin{tabular}{ p{4cm}p{7cm}P{3cm}  }
    \toprule
     Reference & Type of particle & Recommended SOR (m/h)\\
     \midrule
     \citet{Driscoll08book}   &  Mill scale and pulp, paper solids (light particles)    & 0.6 \\
     & Mill scale and pulp, paper solids (heavier ones)  &  2.4 \\
     \citet{Crittenden12book} &  Alumn flocs (\SI{1020}{-} \SI{1100}{kg/m^3}) & 2.5 - 6.5\\
     &   Water treatment (heavier flocs)  & 3.8 - 7.5\\
     \bottomrule
    \end{tabular}
\end{table}

Similar ranges are given elsewhere in the literature~\citep[\eg,][]{Wilson05book,Davis10book}. When microsand particles are added between downstream coagulation and upstream flocculation, SOR values can be enhanced to typical values between \SI{23}{m/h} and \SI{50}{m/h} for clarification in water treatment processes~\citep{Baruth04book}.

\subsection{Settler capacity for monodisperse suspensions}\label{s:settler_capacity}

For suspensions that can be considered in practice as monodisperse, a criterion for the definition of settler capacity is to find the conditions that will ensure that particles will not report to the overflow of the settler. Although the condition of ensuring zero turbidity might be impractical (and should be, in most cases, replaced by the notion of efficiency as described in the next section), the present section uses this condition to develop and put together two possible approaches for capacity estimation of \BES.   %

\subsubsection{Mass balance approach}

\paragraph{Batch case}

In batch (closed) systems, a first explanation of the observed clarification mechanism of inclined tubes compared with vertical ones is attributed to~\citet{Ponder25QJoEPTaI} and~\citet{Nakamura37KJM}. This, so-called PNK theory, states that effect of particle settling beneath an inclined, downward facing wall, is to convey additional clear fluid upwards, above the existing horizontal suspension-supernatant interface. This, by virtue of mass conservation, causes that such suspension-clear fluid interface moves downwards at a faster rate than that occurring in vertical conduits. A geometrical interpretation conducting to the PNK expression (derived hereafter) is reviewed in~\citet{Kinosita49JoCS} and~\citet{Hill77IJoMF}. 

\begin{figure}[!t]
    \centering
    \scalebox{1.25}{\includegraphics{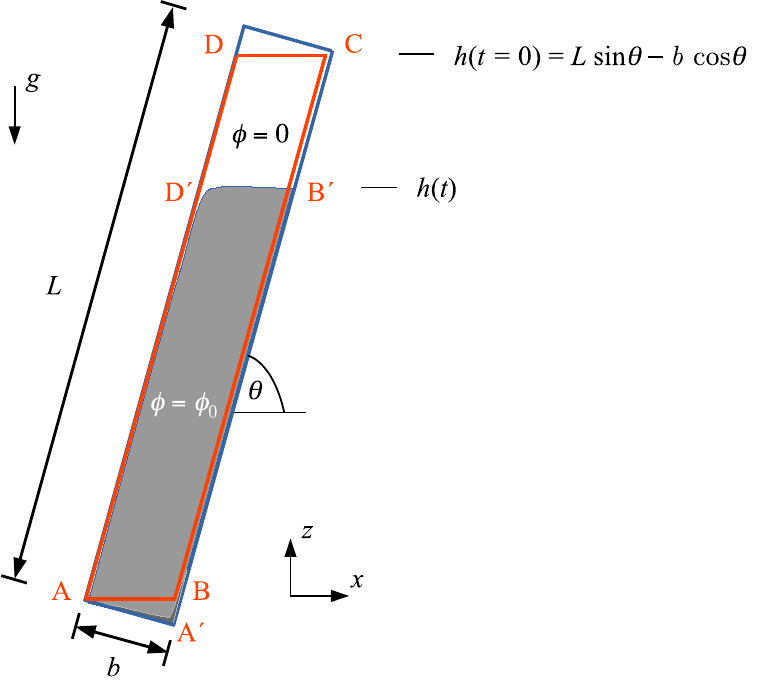}}
    \caption{Schematic of mass balance, \RB{where \textit{g} is the gravity acceleration, \textit{L} is the length of the plate, \textit{b} is the plate spacing, $\theta$ is the tilt angle, \textit{h} is the interface position (height), and $\phi$ is the particle volume fraction.} Note that in this scheme, the distance between the lower plate and the control volume boundary BC, where part of the suspension leaves the control volume, is negligibly small.}
    \label{f:PNK_schematic}
\end{figure}

Consider a tilted, closed rectangular domain as depicted in Figure~\ref{f:PNK_schematic}. Initially, the volume is filled with a monodisperse, homogeneous, and dilute suspension (volume fraction $\Cv_0\ll1$) of spheres of diameter $d_p$ and density $\rhos$, between vertical coordinates $z=-b\cos\tilt$ and $z=\ho$, and where the fluid phase has density $\rhof$ and dynamic viscosity $\muf$. 
Immediately after the start of the experiment, an interface between the suspension and a supernatant layer forms due to the settling process. As the interface descends, its vertical displacement is traced, at each time $t,$ by the vertical level $h(t)$ at which this interface reaches the lower wall, which can in turn be obtained from a solid density function $\tilde{\rho}_s(x,z,t)$ describing the spatial distribution of the suspension, as shown by the gray area in Figure~\ref{f:PNK_schematic}. 
It is assumed that, away from the interface between the suspension and the clear liquid, the solid-phase velocity $\bv{\vset}=\vset\ghat$, where $\ghat$ is the unit vector in the direction of the gravity acceleration, $\bv{g}$, and $\vset>0$. This is the settling velocity $\bv{\vset}$ of the bulk suspension, which depends on the solid and fluid phase properties, the settling element equivalent diameter, and the particle concentration. In the case of a single sphere settling in an otherwise quiescent Newtonian fluid and in the limit of small Reynolds number, then $\vset$ is the Stokes velocity
\begin{equation}\label{e:wSt}
 \wSt= \frac{g(\rhos-\rhof)d_p^2}{18\muf}.   
\end{equation}
More generally, $\vset$ is commonly expressed as $\vset=w_{\text{St}}f(\Cv)$, where $0\leq f\leq1$ and $\ioD{f}{\Cv}<0$~\citep{Davis85ARoFM}. 
A consequence of the dilute hypothesis is that the sediment layer,  where particles accumulate along the lower wall, is very thin and is assumed herein of negligible thickness.
We define the control volume ABCD of Figure~\ref{f:PNK_schematic}, wherein the segment BC is located slightly above the lower wall, say a few particle diameters, just enough so that the negligibly thin sediment layer is left outside of the control volume. 
The opposite segment DA coincides with the location of the upper wall. 

Unless transient behavior occurs when slender containers are used to further accelerate the process~\citep{Herbolzheimer81JoFM}, the interface $\text{B}'\text{D}'$ has been observed to remain mostly horizontal~\citep{Kinosita49JoCS,Acrivos79JoFM}, leaving, as it evolves, a clarified layer above. A density function can be defined within the control volume as~\citep{Ihle21PT}:
\begin{equation}
    \tilde{\rho}_s = 
    \begin{cases}
    \rho_s\Cv_0 & \text{if } (x,z)\in \text{A}\text{B}\text{B}'\text{D}',\\
    0 & \text{otherwise,}
    \end{cases}
\end{equation}
where the time-dependent position of the top of the suspension layer ($\text{B}'\text{D}'$ in Figure~\ref{f:PNK_schematic}), is assumed as a horizontal line at the vertical level $z=h(t).$ 
Differently from the referred, previous geometric interpretations, this specific choice of control volume is independent of the thickness of the clear fluid layer beneath $\text{A}\text{D}'$.
%
Imposing solid phase conservation and using the Reynolds transport theorem ($0=\int_{\Omega}\partial\tilde{\rho}_s/\partial t\dd\Omega + \int_{\partial\Omega}\tilde{\rho}_s\bv{u}_s\cdot\nhat\dd A$, where $\nhat$ is the unit normal pointing outwards
, $\Omega$ and $\partial\Omega$ represent the control volume and its boundary, respectively), the well known PNK expression is obtained: %
\begin{equation}\label{e:pnk-dhdt}
    \oD{h}{t} = -\vset\left(1+\frac{h}{b}\cos\tilt\right),
\end{equation}
where the first and second terms in the right hand side are given by the fluxes through AB and $\text{BB}'$, respectively. 
 
This model has been subject of experimental scrutiny, and differences with the theory have motivated to propose ad-hoc adjustments. \citet{Graham63CJCE} proposed to include a concentration- and geometry-dependent correction factor (between $0.4$ and $0.8$), and their results were sensitive of the inclination angle for values above \SI{70}{\degree} and to particle concentration in the whole range tested. 
The placement of a factor multiplying the whole right hand side of~\eqref{e:pnk-dhdt} caused some scatter in their results. \citet{Vohra71IEC} proposed to modify the location of such correction to solely multiply the term including $h/b$ in~\eqref{e:pnk-dhdt} (linearly increasing with particle concentration and inclination-independent at low concentrations). 
Similarly, for the case of settling in inclined tubes, \citet{Sarmiento78CJoCE} also proposed a correction factor ($F$) to the second term based on the possibility of a horizontal component of the liquid phase flow feeding the clarified layer from the suspension. According to their experimental measurements using lightly flocculated red mud suspensions (\SI{75}{\%} smaller than \SI{6.1}{\micro m}), they proposed a factor of the form $F=k(\phi)/\sin\tilt$, which depends strongly on the suspension characteristics. 
To explain the discrepancies between experimental results and the PNK model~\eqref{e:pnk-dhdt}, \citet{Oliver64CJoCE} relaxed the original mass-balance assumption that the clarified liquid volume below the upper plate reports instantaneously to the overflow. Instead, part of the water in the suspension below the top plate flows into the clear water layer accumulating as it ascends, resulting on a wedge-like clear fluid layer shape. The results of their batch model had best fit at relatively low concentrations and small times; although they report some degree of wave formation, they observed that most of the settling process was in the absence of such disturbances. The discrepancy between the PNK model and experimental results has been explained by~\citet{Zahavi75IECPDaD} by the suspension concentration. In their experimental and theoretical work, they found that enhanced settling in vessels with inclined walls accentuates with concentration (they tested with volume fractions below \SI{10}{\%}) and that discrepancies with the PNK model are partly due to changes in concentration during the settling process, not accounted for in the original model.
A different approach was performed by \citet{Acrivos79JoFM}, who provided a fluid dynamic derivation of the PNK model for containers of general shape and discussed the conditions for its validity, placing the PNK model on a firmer theoretical basis. On the other hand, they explicitly related the PNK model for batch clarifiers to the continuous settling process and highlighted the importance of the stability of the interface as a key limiting factor of the PNK model's applicability. 




%
\paragraph{Continuous feed}
For batch clarifiers, the rate of production of clarified fluid volume (per unit length in the spanwise direction), $\dimClar(t),$ is obtained from the vertical velocity of the interface~\eqref{e:pnk-dhdt} and the horizontal size of the container $b/\sin\tilt$ as $\dimClar(t) = \frac{\vset b}{\sin\tilt}\left(1+\frac{h}{b}\cos\tilt\right)$.
In terms of the more commonly used distance along the lower plate between the edge $\text{A}'$ and the suspension height $\text{B}'$, $l(t)$, given by $l=h/\sin\tilt + b/\tan\tilt$ (Figure~\ref{f:PNK_schematic}), the production rates becomes 
\begin{equation}\label{e:clarification_rate}
    \dimClar(t) = \vset b\left(\frac{l}{b}\cos\tilt+\sin\tilt\right).
\end{equation}


The batch-PNK theory result~\eqref{e:pnk-dhdt} can be extended to countercurrent, steady-state continuous mode of operation, noting that the required feed required to keep the supernatant-suspension boundary at the height corresponding to length $l(t)=L=\ho/\sin\tilt$ (constant) must satisfy $s=\dimClar(l=L)/b\geq \vmean,$ where $\vmean$ is the mean velocity along the settler, and thus~\citep{Davis85ARoFM}:
\begin{equation}\label{e:PNK_vmean}
    \frac{1}{\wparam} \leq \frac{L}{b}\cos\tilt + \sin\tilt, 
\end{equation}
with $\wparam=\vset/\vmean$. The volume flow per unit width $\vset\frac{L}{b}\cos\tilt$ has been identified by~\citet{Herbolzheimer81JoFM} and \citet{Leung83IECPDaD} as (approximately) the maximum theoretical feed capacity of a tube or plate settler in the case when the clear fluid layer is thinner that the suspension layer (see Section~\ref{s:mode_of_operation}, below) or countercurrent flow, and approximately corresponds to~\eqref{e:PNK_vmean} when $\cos\tilt\sim\sin\tilt$ and $L/b\gg1$. It is noted that while in vertical tubes ($\tilt=\SI{90}{\degree}$), the critical particle removal condition is achieved with $\wparam=1$ (\ie the mean flow velocity is equal to the settling velocity), in inclined tubes, depending on $\tilt$ and $L/b$, $\wparam$ could be much larger than the unity (in other words, the settler would separate particles at mean flow velocities largely exceeding the particle settling value), thereby denoting the significant increase of the capacity of the settler that results from inclining conduits.

\RB{According to Equation \eqref{e:clarification_rate}, the clarification rate increases with the aspect ratio $L/b$, but there is a limit regarding real operational conditions that may cause the occlusion of very thin elements.}
Commonly, industrial inclined elements (\eg, tubular) in the wastewater industry feature $L/b$ between close to 50 (see Section~\ref{s:intro}) and larger than 100 in some other applications of mineral processing~\citep{Galvin21CES}. Limitations found in the wastewater industry that deals with flocs\RB{, \eg made of clays or other ultrafine particles,} are the potential of clogging for tubular elements below \SI{40}{mm} in diameter~\citep{Lin14book}, a problem that has also been known for plate-like equipment~\citep{Zhou12AMaM}. \RB{The potential of clogging, has been mostly controlled adjusting the incline angle of cells~\citep[][]{Culp68JAWWA} or exerting vibration either to the cell or to the suspension, as discussed in Section~\ref{s:improvements_settling}.}

In the theoretical analysis made by~\citet{Herbolzheimer81JoFM} and \citet{Leung83IECPDaD}, there is no specific regard to the tubular geometry in their derivation, and the corresponding settling elements used in their experiments were indeed rectangular with the longest side placed vertically. So far the only exceptions using the PNK theory are the derivations in the batch case by~\citet{Graham63CJCE}, who replace $h/b$ in \eqref{e:pnk-dhdt} by $4h/\pi d$, with $d$ the inside diameter of the pipe, and the recent generalization by~\citet{Ihle21PT}, both for batch and continuous geometries, referred in Section~\ref{s:continuous_feed}.

\subsubsection{Particle trajectory approach (continuous feed)}\label{s:continuous_feed}

If particles are smaller than the smallest flow scale and their inertia can be neglected, then a force balance on a single sphere of volume \Vp {}  yields
\begin{equation}\label{e:particle_force_balance}
    \bv{\hat{F}}_{d,p} + V_p\rhos\bv{g} -  V_p\grad p_f = 0,
\end{equation}
where $\bv{\hat{F}}_{d,p}$ is the drag force acting on the particle, and $\grad p_f$ is the local pressure gradient in the fluid phase. The last term comes from the pressure integral $\int_{\partial\Vp}p_f\nhat\,\dd A$ assuming that pressure variations on the particle surface are given by the hydrostatic approximation. Noting that the drag force occurs on $n_p=\Cv/V_p$ particles per unit volume (with $\Cv$ the volume fraction) and that $\bv{F}_{d,f}=-n_p\bv{\hat{F}}_{d,p}$, then
\begin{equation}\label{e:F_df}
\bv{F}_{d,f}=\Cv(\rhos-\rhof)\bv{g},
\end{equation}
where $\bv{F}_{d,f}$ is the volume force density exerted on the fluid phase by the particles. In other words, such volume force corresponds to the immersed weight of the particles. 
An implication of \eqref{e:particle_force_balance} and the hydrostatic assumption is that particles instantaneously adopt a terminal velocity. If particles are spherical and $d$ is their diameter, from the integration of the flow equation around a sphere, in the limit of small particle Reynolds number, the drag force is $\bv{\hat{F}}_{d,p}=-3\pi\muf d(\bv{u}_s-\bv{u}_f)$~\citep[][eq. 126]{Stokes51TCPS}, where $\bv{u}_s$ and $\bv{u}_f$ are the solid and fluid velocities, respectively. Therefore, the solid phase velocity must verify:
\begin{equation}\label{e:rel_vel}
    \bv{u}_s-\bv{u}_f = \wSt\ghat,
\end{equation}
with $\wSt$ given by~\eqref{e:wSt}. Thus, this expression provides a relationship between the absolute velocity of the solid phase, the absolute velocity of the fluid phase, and the settling velocity in the dilute limit.

\begin{figure}[!h]
    \centering
    \scalebox{1.25}{\includegraphics{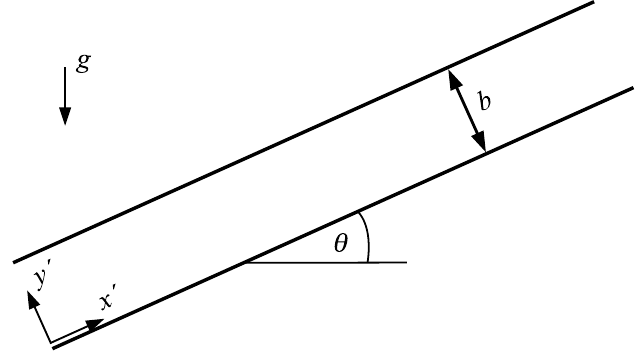}}
    \caption{Schematic of axes in the constant velocity model derivation. The unit vectors associated to the axes $x'$ and $y'$ denoted therein are $\xhat$ and $\yhat$, respectively.}
    \label{f:xy_slm}
\end{figure}

In the wastewater industry, the approach that is most widely referred to for the dimensioning of \BES settling elements~\citep[excluding flow entrance length, as modelled by][]{Fadel90JoEE} is that particles describe a straight-line trajectory starting at any normal coordinate from the bottom plate; this assumes that the system is in the dilute limit, at steady state at each inclined flow conduit, and that the flow is uniform and reasonably well described by the mean flow velocity. 
While \citet{Fadel90JoEE} have argued that the constant velocity model leads to a sub-dimensioning of the tube length, this is still widely recommended in the literature~\citep[\eg,][]{Letterman99book,Wilson05book,Crittenden12book,Lin14book,Droste19book}. If, in this case, particles are nearly spherical (and also sufficiently small to render the particle Reynolds number very small), then the form factor $k=3\pi$ in the Stokes drag force expression above can be replaced by a different one provided that their resistance to the flow is isotropic~\citep{Guazzelli12book}. If so, the Stokes settling velocity in~\eqref{e:rel_vel} can be replaced by $\vset=V_p(\rhos-\rhof)g/kd_p'\mu$ where, in this case, $d_p'$ is an equivalent diameter of the particle, which will depend on its shape~\citep[see, \eg,][]{Komar78JoG,Hartman94IECR,Letterman99book}. Thus, the solid phase velocity can be assumed, using $\bv{\vset}$ in the right hand side of~\eqref{e:rel_vel}, that: 
\begin{equation}
\bv{u}_s = \bv{u}_f + \bv{\vset},
\end{equation}
Defining $\xhat$ and $\yhat$ as unit vectors parallel and normal to the inclined plate (aligned with the axes $x'$ and $y'$ in Figure~\ref{f:xy_slm}), then the solid phase velocity vector corresponds to:
\begin{equation}\label{e:solid_velocity_Culp}
\bv{u}_s=-\left[\vset\sin\tilt + c u_{fx}\right]\xhat + (u_{fy}-\vset\cos\tilt)\yhat,    
\end{equation}
where $c$ is $1$ or $-1$ in cocurrent or countercurrent flow, respectively. Assuming unidirectional fluid phase flow in the direction parallel to the plate, $u_{fy}=0$ and noting that in light of the present approach, the maximum travel time of a single particle, $\tp$, can be computed as that corresponding to a particle departing from the top plate ($y=b$), then~\citep{Letterman99book}:
\begin{equation}\label{e:tper}
    \tp = \frac{b}{\vset\cos\tilt}.
\end{equation}

 \citet{Letterman99book} used directly $\tp$ in~\eqref{e:tper}. Thus, the plate length in excess of the entry value is computed as: 
\begin{equation}\label{e:entrance0}
    L = \frac{\tp}{b}\int_0^b\bv{u}_s\cdot\xhat\dd y,
\end{equation}
where $y$ is the coordinate normal to the plate. Noting that $(1/b)\int_0^b u_{fx}\dd y=\vmean$, the procedure described in~\citet{Letterman99book} suggests that the plate length of the settler (in addition to the entrance length, where the flow is not uniform), can be computed using~\eqref{e:entrance0} as:
\begin{equation}\label{e:Letterman_Lb}
\frac{L}{b}=\frac{\vmean+c \vset\sin\tilt}{\vset\cos\tilt},  
\end{equation}
where the corresponding limiting condition for effective separation of particles is expressed as:
\begin{equation}\label{e:awwa-criterion}
    \frac{1}{\wparam} \leq \frac{L}{b}\cos\tilt {}- c\sin\tilt.
\end{equation}    
It is noted that the countercurrent case ($c=-1$) corresponds to the prediction obtained using the PNK model in \eqref{e:PNK_vmean}. This implies that, for the particular case of the parallel plate geometry, the present approach is accurate from a mass balance perspective. Also, in the case of the countercurrent case, from~\eqref{e:Letterman_Lb}, it is concluded that $(L/b)\cos\tilt<1/\wparam$, which thereby confirms the maximum throughput conclusion obtained using boundary layer flow arguments~\citep{Herbolzheimer81JoFM,Leung83IECPDaD}.

\begin{figure}[!h]
    \centering
    \scalebox{1.25}{\includegraphics{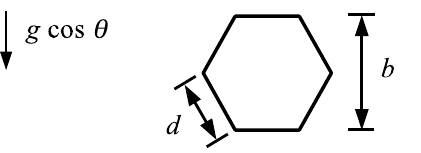}}
    \caption{Regular hexagon of side $d$ and height $b$.}
    \label{f:hex-pack}
\end{figure}

This simplified approach can be applied to a different (constant) cross-sectional area, provided $b$ in~\eqref{e:tper} is replaced by the maximum top-to-bottom distance normal to the flow plane, thereby representing the largest possible travel time of a single particle. For instance, in the case of a pack of regular hexagons of side $d$ (Figure~\ref{f:hex-pack}), the largest travel distance of particles, denoted as the perpendicular line between the top plane and the bottom plane in the figure, is given by $b=d\sqrt{3}$. In this example, if the volume flow at each conduit ($q$) is known, then the mean velocity can be computed, in the case of the hexagonal cells, as $\vmean=\frac{2q}{3\sqrt{3}d^2}$.

Key to the derivation of the criterion~\eqref{e:awwa-criterion} is that particles move along the channel with characteristic longitudinal velocity $\vmean$ corresponding to the mean flow velocity. \citet{Yao70JotWPCF} relaxed this restriction including in his analysis the vertical slice of the velocity profile corresponding to the longest vertical settling section (\eg, a central vertical plane in pipe flow), and proceeded with integration to find a condition to ensure that particles will settle in the conduit rather than report to the overflow of the clarifier:
\begin{equation}\label{e:Yao70}
    \frac{\SY_c}{\wparam} < \frac{L}{b}\cos\tilt + \sin\tilt
\end{equation}
where $b$ shares the same definition as in~\eqref{e:awwa-criterion} and $\SY_c$ is a shape factor that results from imposing the boundary conditions of the problem (\ie, at the top and the bottom of the flow slice of interest). Although his calculations were done only for the countercurrent condition, it is possible to formulate them also for the cocurrent case, whereby an analogous expression to~\eqref{e:awwa-criterion} can be found. The parameter $\SY_c$ is defined as:
\begin{equation}\label{e:SYc}
\SY_c=\frac{1}{b\vmean}\int_0^b u_{y,\text{cp}}\dd y,    
\end{equation}
where $u_{y,\text{cp}}$ is the (permanent, laminar, and independent of the longitudinal coordinate) velocity profile of the vertical slice of height $b$, which is that associated with the maximum particle travel distance. For $(L/b)\cos\tilt + \sin\tilt> \SY_c/\wparam$, particles will not report to the overflow. \citet{Yao70JotWPCF} computed $\SY_c=1$ for the parallel plate geometry, $4/3$ for tubes, and $11/8$ for square conduits. This widely cited approach has also been subject to criticism, being overly conservative~\citep{Fadel90JoEE}. 
Recently, combining mass balance and the concept of falling distance along the conduit resulting in~\eqref{e:SYc}, \citet{Ihle21PT} proposed a new, less conservative criterion that extends the set of cross-sectional geometries originally reported by \citet{Yao70JotWPCF} and provides additional values of $\SY_c$.

Other approaches in the design of settlers have been discussed by~\citet{Willis78JAWWA}, \citet{Ziolo96CES}, and \citet{Bandrowski97CEaPPI}, including asymptotic approaches (\ie, those predicting complete clarification only in the limit $L/b\to\infty$) and a polydisperse formulation that starts from a particle particle size distribution representing that of the suspension, which is used to define a settling velocity distribution~\citep[a version of this model, adapted to a lamella settler pack fitted as a settling ring at the overflow of a vertical thickener is proposed by][]{Kowalski04AoHEaEM}.
%
%
After a comparison and validation campaign using two independent datasets, \citet{Bandrowski97CEaPPI}, concluded that the model based on a single cutoff settling velocity including~\eqref{e:awwa-criterion} (countercurrent flow) and~\eqref{e:Yao70} are of comparable accuracy than his asymptotic model and had less deviations from experimental data than others~\citep[described by][]{Ziolo96CES} and can yield similar results than those including size distributions.

\subsection{Efficiency}\label{s:efficiency}

In most industrial processes, it is virtually impossible to obtain a particle-free overflow due to equipment and suspension conditions and yet some degree of turbidity is acceptable. In clarifiers with dilute feed, the most widespread process indicator of efficiency in this context corresponds to the \emph{turbidity removal efficiency} (\TRE):
\begin{equation}\label{e:TRE}
    \TRE = \frac{\turb_0-\turb_{\OF}}{\turb_{0}},
\end{equation}
where $\turb_0$ and $\turb_{\OF}$ represent the turbidity at the feed and overflow, respectively~\citep[typically measured in NTU and commonly linearly related with particle concentration with high correlation coefficients, as described by][and references therein]{Bertrand04WSaT,Saady12WSaT}. Fixing the clarifier and the suspension characteristics, the \TRE is most commonly a decreasing function of the feed flow and thus of the SOR. The experimental study of~\citet{Jimenez97ET} with agroindustrial biological flocs shows a linear SOR-\TRE behavior both using plates and tubes as \BES elements, with \TRE-values above \SI{80}{\%}. For storm-water turbidity removal purposes, \citet{Clark09JoEE} report composite data revealing a drop in $d_{50}$ from \SI{70}{\micro m}--\SI{220}{\micro m} to \ca \SI{30}{\micro m}, associated with a SOR range between \SI{1.44}{m/h} and \SI{7.2}{m/h} and \TRE-values between \SI{64}{\%} and \SI{98}{\%}. Using in their experiments sand and ground silica, they found a tendency to a decreasing  SOR-\TRE curve with a stronger scattering in the low flow end. 
\citet{Salem11WR} made experiments using walnut shell particles (density about half of that of natural sediments). Instead of monitoring turbidity, they measured solid concentrations at the feed and the overflow and used a definition of separation efficiency (SE) analogous to~\eqref{e:TRE}, finding a relatively low-efficiency range between \ca~\SI{27}{\%} and \SI{45}{\%} testing three designs of settler feeds. In their flow rate-SE determination, they found a tendency to a lower plateau for relatively high feed flows. Although they did not report the flow Reynolds number within the settling elements they used, it is apparent that such low SE-values were found because they configured the flow in their experiments either turbulent or transitional (at least, that was their numerical modeling hypothesis).
\citet{Saady12WSaT} studied the problem of clarification of river water with very fine particle contents, an application that resembles the problem of thickener overflow cleaning under the presence of fine particles. Using alum (aluminum sulfate) coagulants, he produced flocs of size distribution finer than \ca \SI{30}{\micro m} and reported turbidity removal efficiencies between about \SI{83}{\%} and \SI{95}{\%} for SOR values between \SI{1}{m/h} and \SI{3}{m/h}, fitting logarithmically-decreasing SOR-\TRE curves \citep[in a previous contribution, the same author fitted \TRE curves to $\TRE\propto \text{SOR}^{-a}$, with $a\approx -0.1$][]{}.

\RB{Table \ref{t:Efficiency} presents a comparison of the main parameters of the above discussed reports from previous authors.} 



\begin{table}[!h]
\centering
\caption{Reported efficiency for SIS processes.}
\label{t:Efficiency}
    \begin{tabular}{ p{4cm}p{3cm}p{4cm}P{3cm}  }
    \toprule
     Reference & Type of floc & SOR-TRE curve behaviour & Efficiency (TRE \%)\\
     \midrule
     \citet{Jimenez97ET}   & Agroindustrial biological flocs    & Linear & >80\\
     \citet{Clark09JoEE} & Storm water turbidity removal  &  Decreasing, strong scattering in the lower flow end & 64-98 \\
     \citet{Salem11WR} &  Walnut shell particles & Tendency to a lower pateau for high feed flows & 24-45\\
     \citet{Saady12WSaT} &   River water with fine particles  & Logarithmically decreasing & 83-95\\
     \bottomrule
    \end{tabular}
\end{table}

An alternative means of assessing the performance of the \BES system concerning process characteristics (\ie, how much of the feed can actually separate, which is measured using the \TRE parameter) is to compare it with the maximum possible value. This corresponds to the PNK result, which, as has been pointed out, has been noted theoretically to be the maximum attainable throughput~\citep{Leung83IECPDaD,Davis85ARoFM} given system geometry and suspension characteristics. \citet{Borhan89PoF} has defined such overall efficiency as:
\begin{equation}
    \EG = \frac{\qmax}{\qPNK},
\end{equation}
where $\qmax$ is the maximum possible flow that yields zero turbidity at the overflow, and $\qPNK=\vmean b$ is the feed volume flow (in this case, per unit width) obtained from~\eqref{e:PNK_vmean}. As the settling velocity ($\vset$) includes, from an empirical perspective, the effect of particle shape and suspension concentration, the only factor that can render $\EG<1$ is the development of flow instabilities within the cell. In his experiment on countercurrent (subcritical flow) flow, \citet{Borhan88PoF} found  higher values of $\EG$ after increasing particle concentrations. This was, however, at a cost of clarification rate ($\dimClar$), given the reduction of the settling rate due to such increase of concentration. Also, similarly as in~\citet{Leung83IECPDaD}, he found that higher efficiencies are found at lower inclination angles.



\subsection{\RB{Connection between SOR and overflow turbidity}}

Given a settler geometry and suspension characteristics, it has been mentioned above that it is possible that the SOR can be related to the \TRE, in a way that increasing overflow rates comes at a cost of increased turbidity. There is, however, a single theoretical maximum value of the surface overflow rate below which there is not particle flow at the overflow (\ie $\TRE=1$), and corresponds to the clarification rate $\text{SOR}/n=\dimClar(l=L)/nb$, where $\dimClar$ is given by~\eqref{e:clarification_rate} and $n$ is the number of settling elements of the clarifier. Thus:
\begin{equation}\label{e:SOR_w}
    \max_{\TRE=1}\text{SOR} = \vset\left(\frac{L}{b}\cos\tilt + \sin\tilt\right).
\end{equation}
In other words, the limiting condition to avoid particle presence at the overflow depends on the settling velocity, which can be regarded as an intrinsic property of the suspension. 


\section{\RB{Flow dynamics in \BES}}\label{s:sedimentingFlow}

The research work on the flow dynamics reported during the second half of the 1970s~\citep[probably starting with][]{Hill77IJoMF,Probstein77proc} represents an inflection point on the understanding of the underlying physics behind \BES technology and gave birth to a modern approach of modeling this kind of equipment.

Using the dilute limit approximation, \citet{Hill77IJoMF} used the same velocity and time scales indicated herein and proposed a set of mixture equations to numerically describe the flow under a cone. 
This pioneering work concurred with the layered model proposed by~\citet{Probstein77proc}, also based on the dynamic balance in the flow as opposed to the previous kinematic approaches that attempted to correct~\eqref{e:pnk-dhdt} to describe more realistic conditions~\citep[see][for early references in the subject]{Oliver64CJoCE,Hill77IJoMF,Acrivos79JoFM}. In particular, \citet{Hill77IJoMF} identified a dependence of the flow features with a Grashof and a Reynolds number ($\Gr$ and $\NRe$, respectively), defined as:
\begin{equation}\label{e:Ge_Re}
    \Gr = \frac{\rhof g \Cv_0(\rhos-\rhof)\ho^3}{\muf^2};\quad 
    \NRe= \frac{\rhof g (\rhos-\rhof)d^2\ho}{18\muf^2},
\end{equation}
and, moreover, the ratio
\begin{equation}\label{e:rGR}
    \rGR = \frac{\Gr}{\NRe}.
\end{equation}

The Grashof number represents a balance between buoyancy (here, induced by particle settling) and viscous dissipation force, and the Reynolds number a balance between inertial and viscous force. Therefore, $\rGR$ represents a balance between buoyancy and inertial force. It is noted that industrial-sized equipment for solid-liquid separation using inclined plates have values of $\rGR$ on the order of $\num{e6}$~\citep{Leung83IECPDaD}. This confirms the pertinence of the boundary layer approximation. The first research that sought to apply the boundary layer concept to describe the details of settling in inclined containers was due to~\citet{Acrivos79JoFM}. Starting from a dynamic formulation similar to that of~\citet{Hill77IJoMF} (albeit, correcting both dynamic viscosity and settling velocity with the average effect of particles), they derived and solved a boundary value problem similar to the Falkner-Skan equation~\citep{Currie12book}. They also showed from their fluid dynamic approach that the mass balance statement~\eqref{e:pnk-dhdt} is valid in the limit of large values of $\rGR^{1/3}$. Their approach, centered on the clarified and the suspension layer was complemented by \citet{Schneider82JoFM} with a drift flux model that identifies feasible suspension-sediment layer combinations according to the suspension concentration-dependent response, by analogy with the Kynch theory of (vertical) thickener-clarifiers, discussed in detail by~\citet{Concha14book}.

For the case of slender containers, \citet{Herbolzheimer81JoFM} and~\citet{Leung83IECPDaD} provide expressions to estimate thicknesses in cocurrent and countercurrent systems. It is noted that, when it evolves very thin compared to the plate spacing and prevails a balance between buoyancy and viscous forces, the clear fluid layer thickness away from the bottom edge of the countercurrent settler is approximately proportional to $x^{1/3}$, where $x$ is the longitudinal coordinate  measured from the upper plate of the bottom edge~\citep[][provided a numerical verification of this behavior]{Laux97PT}. A derivation of this exponent using boundary layer arguments is given in~\ref{a:scaling-derivation}. 

In countercurrent flow, a shear stress on top of the sediment layer exerted by the feed (carrying the suspension) is added against the sediment flow direction, which makes common that higher angles than in cocurrent equipment are required to maximize performance~\citep{Leung83IECPDaD}.


The internal flow structure of \BES cells can be summarized conceptually on three layers: (a) A clear layer near the top where all or part of it may report to the overflow, (b) a suspension layer of particle volume fraction equal to that of the feed ($\phi_0$) that may flow either upwards or downwards, and (c) a sludge layer, flowing downwards, of volume fraction $\phi_U$, that results from particle settling within the cell. 

The angle of the settler should be sufficient to allow for the sludge to continuously flow towards the underflow. Such condition is the result of complex inter-particle interactions that can be interpreted as an equivalent non-Newtonian fluid, not only in the case of suspensions prone to the effect of physico-chemical properties~\citep{Seyssiecq03BEJ}, but also in the simplest, non-cohesive case, defined solely by hydrodynamic forces. While early interpretations of these dense currents modelled particle flow as the result of a diffusion-like process~\citep{Kapoor95JoFM}, a more recent approach embrace equivalent rheological properties in terms of a viscous number, defined as a balance between particle a rearrangement time scale and a strain one, where the relevance of particle inertia is a defining element~\citep{Boyer11PRL}. Examples of slowly flowing, non-colloidal particle gravity currents, interpreted in terms of such viscous number can be found both with negligible confinement in published experimental work~\citep{Cassar05PoF,Palma16PoF} and with strong confinement in the numerical simulations of~\citet{Palma18PT}.
\begin{figure}
    \centering
   \begin{tabular}{cc}
        (a) Cocurrent subcritical & (b) Cocurrent supercritical \\
        \includegraphics[scale=1]{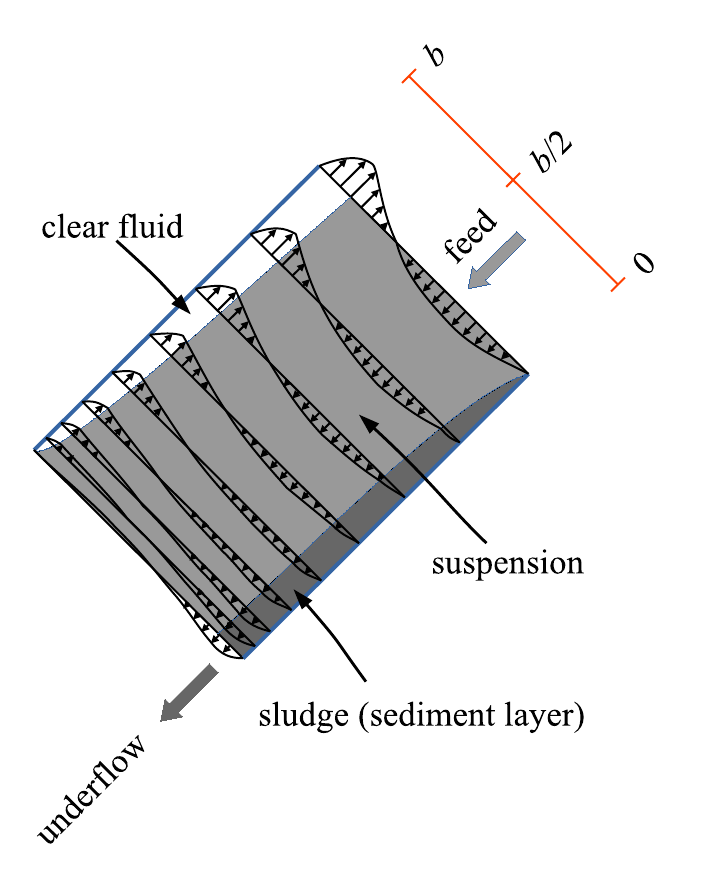} & 
        \includegraphics[scale=1]{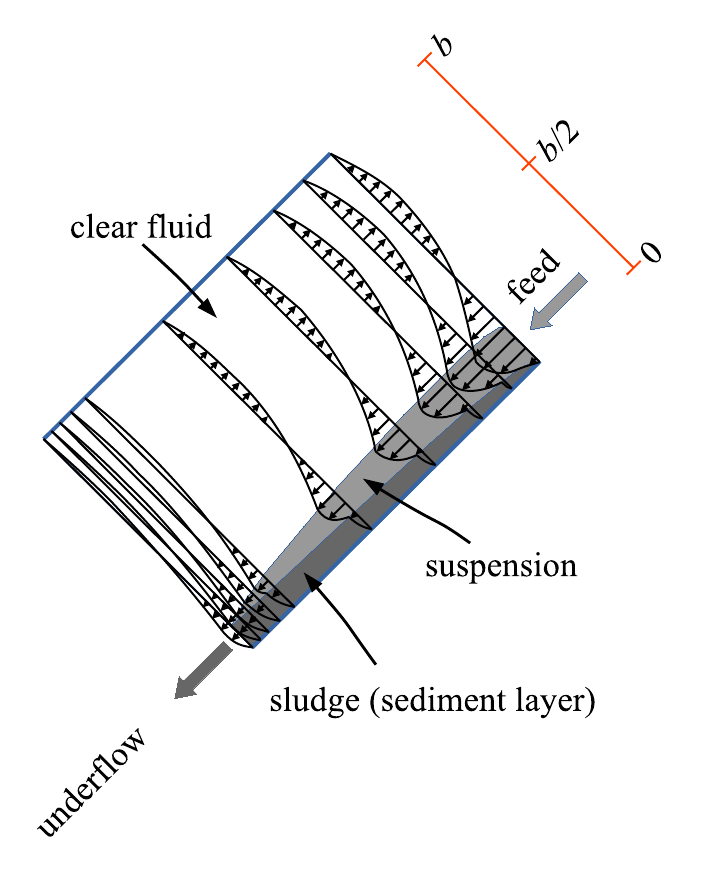}\\
        \multicolumn{2}{c}{(c) Countercurrent (subcritical)}\\
        \multicolumn{2}{c}{%
        \includegraphics[scale=1]{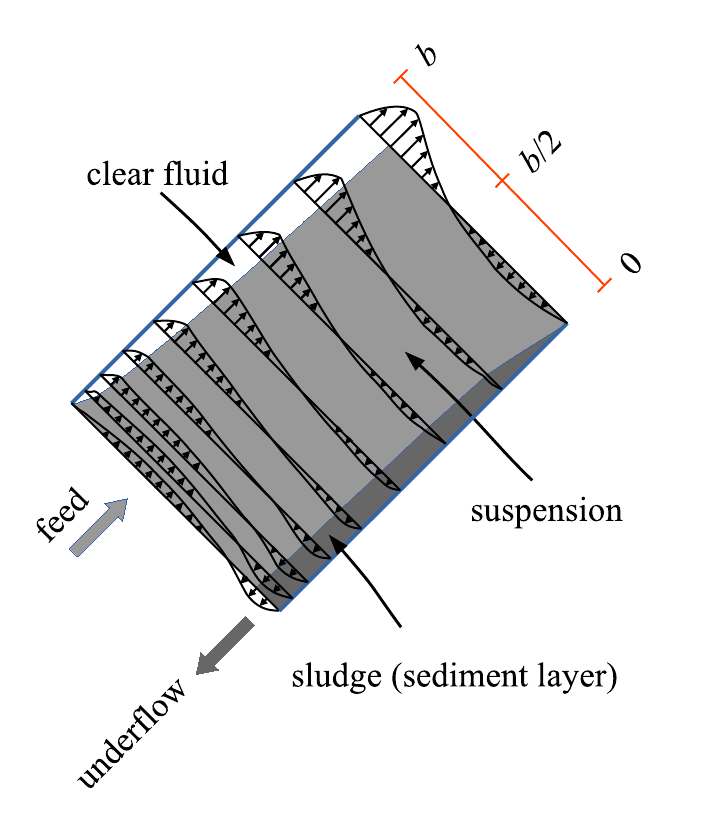}%
        }
    \end{tabular}
    \caption{Schematic of subcritical and supercritical flow for cocurrent operation ---(a) and (b)--- and countercurrent operation (c). Adapted from~\citet{Leung83IECPDaD}.}
    \label{f:subcritical-supercritical}
\end{figure}

The first non-uniform flow model to estimate the capacity and essential flow features of \BES cells has been reported by~\citet{Probstein77proc}. In their work, they formulated three (vectorial) momentum equations associated to a clarified, a suspension and a sludge layer, which have been connected, besides setting top and bottom kinematic boundary conditions, through imposing velocity and stress continuity, shared longitudinal values of pressure gradient in all three layers and the account for inter-layer solid-phase conservation. Additionally, they assumed that the suspension viscosity is the same as that of the clarified layer and lower than that of the sludge layer. Solving for the particular case of zero net volume flow across any particular plane normal to the settling element, they obtained piecewise parabolic velocity profiles and found that the maximum theoretical capacity should take place when the clarified layer width is half of the width of the spacing between the plates.

\subsection{Mode of operation (flow configuration)}\label{s:mode_of_operation}

The aforementioned maximum capacity, given by a clear layer of about half width of the channel, gained significance in distinguishing two different modes of operation for a given total flow rate: the so-called subcritical and supercritical flows depending on whether the clear fluid layer is thinner or larger than the nearly half width that maximizes capacity. 
The model of~\citet{Probstein77proc} was complemented by~\citet{Rubinstein80IJoMF} and \citet{Herbolzheimer81JoFM} using boundary layer concepts, and further continued by \citet{Leung83IECPDaD} to include both cocurrent and countercurrent flow configurations, thereby relaxing their original zero-net-flux assumption. 

In steady-state operation, the clear fluid layer may contract or expand moving away from the feed as it occupies less than half of the space between the plates (Figures~\ref{f:subcritical-supercritical}a and \ref{f:subcritical-supercritical}c, respectively), thereby defining subcritical flow\footnote{This excludes the flow near the feed and outlet, where inertia is no longer negligible even in laminar flow.}, or expand as it moves away from the feed while it occupies more than half of the plate (Figure~\ref{f:subcritical-supercritical}b), thus defining supercritical flow~\citep{Rubinstein80IJoMF,Leung83IECPDaD}. Depending on the configuration of the settler, this might imply, as a result of mass and momentum balance into the system, a flow reversal of part of the feed in subcritical flow~\citep[Figures~\ref{f:subcritical-supercritical}a and \ref{f:subcritical-supercritical}c, based on results obtained by][]{Leung83IECPDaD}.

\citet{Leung83IECPDaD} showed both theoretically and experimentally (using top- and middle-feed points in an inclined cell) that the cocurrent settler can withstand the subcritical and the supercritical configuration, thus confirming previous theoretical predictions~\citep{Probstein77proc,Herbolzheimer81JoFM}. In their experimental apparatus, they were able to trigger the various possible flow conditions by restricting the width of the feed relative to the plate spacing using an adjustable flipper. They also confirmed the previous finding of \citet{Probstein77proc} that a clarified layer thickness greater than half of the plate spacing results in streamwise contracting flow and, conversely, that clear fluid thicknesses of less than \ca $b/2$~\citep[the exact value depending on $\phi_0/\phi_U$, with $\phi_0$ and $\phi_U$ the feed and underflow volume fraction, respectively, as described in][]{Leung83IECPDaD} corresponds to streamwise expanding flow.

According to \citet{Leung83IECPDaD}, the countercurrent mode has only one feasible steady solution in subcritical flow; they showed that, although there is a mathematical solution for the supercritical flow branch, it is unfeasible because it violates mass conservation in the system. 
To the best knowledge of the authors, there is no experimental evidence of supercritical flows in countercurrent settlers. 






\citet{Probstein77proc} and~\citet{Leung83IECPDaD} have argued that in the supercritical mode of operation (attainable in the cocurrent configuration), comparatively higher throughput at equal geometrical and suspension configurations can be achieved prior to the development of instabilities compared with the subcritical mode (indeed, in subcritical flow, part of the feed tends to reverse, as seen in Figures~\ref{f:subcritical-supercritical}a and~\ref{f:subcritical-supercritical}c). However, the cocurrent configuration has inherent limitations to the maximum aspect ratio to achieve steady-state and thereby has a narrower steady operational range. In contrast, in the case of the countercurrent configuration, it is comparatively easier to increase the throughput by enlarging the aspect ratio of the settling elements~\citep[\ie, increasing $L/b$,][]{Herbolzheimer81JoFM,Davis83PoF}. 
In all cases, the development of instabilities plays an important role in the performance of \BES.

\subsection{Flow instability}

To enable an adequate separation, \BES operation requires laminar flow to occur, thereby avoiding mixing between the suspension and the clarified fluid layer. 
In the present context, flow instability corresponds to the departure from laminar flow away from the inlet and outlet zones of the cell, where flow is likely to have a non-negligible inertial component. In other words, the flow in the interior of the cell becomes unstable when flow inertia outstrips viscous forces in such a way that two- or three-dimensional flow structures can hinder the settling process by lifting particles with the potential to eventually carry them to the overflow of the settler. This mechanism is responsible for detriments to the $\TRE$.


It is apparent from~\eqref{e:SOR_definition} that the SOR can be increased by increasing the inclination angle, $\tilt$. However, increasing the SOR for a fixed settling cell causes a decrease of the \TRE, because a cell with a lower retention capacity (\ie, available projected area for settling, $nA_E$) that is receiving the same feed flow will be more likely to report more particles to the overflow. 
On the other hand, not only the settling velocity but the stability of the whole settling system is also dependent on $\tilt$.
%
Looking for best operational values of the latter parameter, \citet{Culp68JAWWA} and \citet{Jimenez97ET} found optimal angles for wastewater treatment in the $\tilt$-range~\SI{35}{\degree}--\SI{45}{\degree}. Mixing bentonite with alum as a coagulant, \citet{Demir95WR} found an overall best among various SOR values around $\tilt=\SI{50}{\degree}$, a range that was later confirmed with the set of high-resolution numerical simulations by~\citet{Chang19PT}.

The fact that the present flow is laden with particles has implications on the interaction between the clarified and the suspension layer, and it is thus expected that the flow will become unstable under conditions that differ from those corresponding to the homogeneous, angle-independent configuration that corresponds to inclined pipes carrying water. Although the result~\eqref{e:PNK_vmean} is independent of the details of the flow, it requires steady flow and the development of a clear fluid boundary layer in the absence of wave formation that may subsequently break~\citep[abundant experimental evidence of wave formation exists in the literature, including][]{Kinosita49JoCS,Oliver64CJoCE,Zahavi75IECPDaD,Probstein77proc,Davis83PoF,Leung83b_IECPDaD,Herbolzheimer83PoF,Brown86thesis,Shaqfeh87PoFa,Borhan89PoF}. While the conditions for steady-state solutions will depend both on the flow regime (subcritical or supercritical) and the operational mode (cocurrent or countercurrent), the onset of such instabilities will also depend on the combination of external variables such as conduit length, inclination, and characteristic clearance, fluid properties and the internal balance between particle buoyancy, flow inertia and viscous forces~\citep{Leung83b_IECPDaD,Shaqfeh86PoF}, where the latter depends on the effect of the particle ensemble on the bulk suspension dynamics~\citep{Borhan88PoF,Borhan89PoF}.

\citet{Leung83b_IECPDaD} analyzed an ideal cocurrent system. For $0 < \tilt < \SI{10}{\degree}$ measured from the horizontal, the onset of waves that break at their crest forming cusps due to a shear destabilization mechanism is reported. This instability mode is reminiscent of the Holmboe modes observed in the inclined, single-phase density stratified tube by \citet{Meyer14JoFM}, albeit in the case of \citet{Leung83b_IECPDaD} findings, symmetry is lost on the suspension side, potentially due to additional energy dissipation resulting from the interaction between particles and fluid. In this small-angle case, little mixing between the suspension and the clear fluid layer was found after cusp appearance, and critical Reynolds numbers within this angle range were weakly dependent on $\tilt$. For higher angles ($\SI{10}{\degree}\leq \tilt\leq \SI{60}{\degree}$) the author describes waves developing and breaking with stronger mixing. Notably, \citet{Leung83b_IECPDaD} identified the same gravity mechanism that triggers roll waves of laminar flows as firstly described by~\citet{Benjamin57JoFM} \citep[and soon after][]{Yih63PoF}, who found spatially developing long waves disturbances around a viscous base state. In this case, the system becomes progressively less stable, due to gravity, with increasing $\tilt$. On the other hand, mixing has been noted to intensify, in both cases, in the downstream direction, as waves develop. \citet{Davis83PoF} made a linear stability analysis in slender, countercurrent (subcritical) continuous flow. At equal geometry, they obtained the same trend, which is also found for particle concentrations above the dilute limit~\citep{Borhan88PoF,Borhan89PoF}, and a considerable stabilizing effect for particle concentrations by volume up to \SI{25}{\%}~\citep{Borhan89PoF}. 


In the subcritical mode, both the clearance of settling elements (\eg, their hydraulic diameter) and their length are key to control the onset and development of waves. In their stability analysis, \citet{Davis83PoF} used the control parameter (written in terms of the present variable definition) $\ReCell=\rhof x \vset/\muf$, where $x\in[0,L]$ is a longitudinal coordinate measured from the bottom of the settling element. After solving an eigenvalue problem, they found marginal stability curves of the parameter $\ReCell\cos\tilt$ as a function of $x/L$ for various values of a dimensionless parameter equal $\ba\equiv\rGR^{1/3}\csc(\tilt) b/L$, with $\rGR$ defined in~\eqref{e:rGR}, and fixed values of $\tilt$. The minima of such curves ($\min_{x/L}\ReCell(x/L,\ba)\equiv\ReCellb(\ba)$), which corresponds to the least stable point for fixed $\ba$ and $\tilt$, is a decreasing function of $\ba$. In other words, as anticipated above, a means to avoid the onset of mixing due to the development of such longitudinal waves in long tubes is to make them slender enough, as also confirmed by the experiments of~\citet{Borhan89PoF}. The definition of $\ReCell$ shows that it can be decreased below $\ReCellb$ by increasing the fluid viscosity and keeping the rest of the system parameters constant, thus allowing the system to avoid interfacial waves with subsequent breakup. 

Theoretically, the aforementioned stability studies~\citep{Herbolzheimer83PoF,Davis83PoF,Shaqfeh86PoF,Borhan88PoF} are spatial stability analyses, which are based on a parallel flow assumption and focus on the growth rate in space of disturbances which are periodic in time. More recent techniques can allow a better understanding and more complete characterization of these instabilities, including the passage from local to global analysis~\citep{Chomaz05ARFM} and the determination of the most dangerous perturbations~\citep{Schmid07ARFM}. The latter are `optimal perturbations', which provide a rigorous bound on the growth of potentially harmful disturbances, and can be computed on time dependent base flows~\citep{Arratia13JoFM} and include nonlinear effects~\citep{Kerswell18ARFM}.

With the present, relatively affordable access to computational resources for massive computational fluid dynamics (CFD) simulations, the linear stability analysis approach finds a significant complement in the study of the nonlinear and potentially three-dimensional development of the flow occurring in particle laden inclined conduits. 
The potential of numerical methods to reproduce flow instabilities has been established with different approaches. 
\citet{Laux97PT} used, for the first time, an Eulerian-Eulerian two-phase finite volume model where they compared their calculations with the PNK model and their own experiments. They observed that while the PNK theory represents well the dynamics of the system, an instability near the top of the suspension was found, resulting on a diffuse suspension-supernatant layer interface. Increasing the particle concentration, they found a departure between their results and the PNK model~\citep[as][also observed]{Borhan89PoF}.
Using a multi-phase particle-in-cell method, consisting of a Eulerian-Lagrangian method for the fluid and particle phase, respectively, \citet{Snider98IJoMF} were able to reproduce numerically cases with and without interfacial waves previously observed by \citet{Herbolzheimer83PoF} \citep[see also][for a validated variation of the numerical method]{Patankar01IJoMPF}. Using a Lattice Boltzmann approach, in a batch case example with $\tilt=\SI{70}{\degree}$ with $\Cv=\SI{10}{\%}$ and particle Reynolds number equal $3.3$, \citet{Xu05CEC} found the existence of local circulations strong enough to sweep particles from central positions towards the boundaries of the settling element, thereby causing some particles to move upwards. As expected from these observations, their results did not match the PNK model.

Studying multiple cell systems, \citet{Nguyen19JoWaET} used a two-phase $k$-$\epsilon$ turbulence closure in a two-dimensional geometry, showing that vortex zones of a settling tank fitted with an inclined plate array are reduced to a half when compared to a tank without inclined plates, both in between and below them, thus reducing overall particle resuspension. In addition, they found that reducing plate spacing enhances particle retention in the plate zone, thus increasing $\TRE$. However, they argued that such a performance improvement depends on the particle size rating, which is reasonable given that particle size controls the settling velocity $\vset$, and thus the Boycott effect, but too large values of the particle size would not be related to a significant effect of plates.
Finally, \citet{Chang19PT} made three-dimensional numerical simulations of a non-dilute ($\Cv_0=\SI{10}{\%}$) batch settling process of a nearly square section domain, with particles between \SI{100}{\micro m}--\SI{150}{\micro m} and density \SI{2650}{kg/m^3}. The resulting dynamics of their simulations was that of a buoyancy-inertia-dominated flow~\citep[between 1 and 2 orders of magnitude higher than the analysis by][]{Davis83PoF} that rapidly evolved to turbulent flow in most of their test cases. Their results showed a tendency to the onset of flow instabilities for $\tilt<\SI{45}{\degree}$ and a decisive role of the particle size for the case $\tilt=\SI{55}{\degree}$ (they found that the system was stable with \SI{150}{\micro m} particles but unstable with \SI{100}{\micro m} ones). In light of their results, the authors suggest that disturbances behave in a quasi two-dimensional fashion. 

%

The aforementioned combination of longitudinal and cross-sectional geometrical characteristics, as well as the fluctuations of the suspension properties and limitations on how thin can settling elements be, makes it a challenging issue to determine the stability of \BES operation. Nonetheless, the dimensionless parameters obtained from the hydrodynamic stability formulations referred to above seem to give a more detailed prediction of conditions for the onset of particle resuspension than those adopted in common water treatment-oriented design approaches, which consider pipe Reynolds numbers within the laminar flow regime (defined as $\NRed=\rhof\vmean d_H/\muf$, with $d_H$ the hydraulic diameter of the system), normally fixed reference values of tube lengths regardless the rest of process and suspension conditions, and minimum diameters selected to avoid clogging~\citep[\eg][]{Letterman99book,Lin14book}, but not necessarily to maximize \TRE. In the context of thickener overflow water treatment in mineral processing plants, there is considerable room for optimized design approaches in this regard, as discussed in Section~\ref{s:BES_technology}.

\section{Advances in \BES-related technology}\label{s:BES_technology}

\subsection{Improvements of settling element array configuration}\label{s:improvements_settling}

Combinations of the cocurrent-countercurrent modes of operation have been developed either by connecting both kinds of equipment through the sludge collection box~\citep[\ie, in parallel, as described in][and references therein]{Brown86thesis} or by placing the feed below the countercurrent operation and above the cocurrent one~\citep{Brown86thesis,Nguyentranlam04IJoMP}. 
The latter configuration constitutes part of the essence of the Reflux Classifier, an array of cocurrent and countercurrent (with the possibility to combine with centrifugal force), closely spaced plate settlers, that is used for elutriation and flotation~\citep{Nguyentranlam04IJoMP,Galvin21CES}. 

The invention of the Reflux Classifier (RC) constitutes the major breakthrough in the application of \BES technology in mineral processing. 
Its development has been extensively reported, and its evolution has been summarized in recent years~\citep{Tripathy15PT}. The performance of arrays of \BES units has been tested either by recycling part of the underflow~\citep{Zhang90IECR} or by sequential connections of inclined tube settlers~\citep[][and references therein]{Saady12WSaT}, in both cases with increased performance when compared with those of single equipment.
Current developments of the RC feature plate spacings on the order of the millimeters~\citep{Galvin21CES}, thus confirming the theoretical enhancement of settling that results from increasing the ratio $L/b$~\citep[in][an example with $L/b>500$ is referred]{Galvin21CES}. In principle, there is no limitation to use this technology for clarification purposes. 

A potential constructive improvement to otherwise vertical, rectangular tanks, has been proposed and extensively tested by~\citet{Fujisaki10WSaT}, who added to the concept of a packed, vertical array of plates with an overflow pipe connection to group them, where overflow collection could be done in combination with pumping suction from downstrean the system. In such work, the only plate spacing analyzed was \SI{50}{mm}, tighter clearances could be tested as a means to increase the performance of the units. Another potential improvement would be testing the concept at higher concentrations.

A variation of \BES technology to improve the performance of units is to include vibration in settler modules, which is an extension of the previous idea of using vibration to mobilize the sludge from the underflow box~\citep{Brown86thesis}. Besides reviewing some advances and applications before 2010, \citet[][and references therein]{Zhou12AMaM} focuses on the use of vibration as a means to avoid clogging in confined settling elements. The authors, who targeted their application to mineral processing, propose the use of inclined plate modules (\ie, a single row of parallel, inclined plates) that feature a shaker placed on the top of the module and a set of vertically-oriented springs to prevent excessive vibration \RB{transmitting} to the structure. In their work, they neither disclose shaking amplitude nor rotation frequency. The same type of enhancement~\citep[also based on previous work by][for cell retention purposes]{Searles94BP} was applied in an algal sediment flow system~\citep{Smith13CES} laterally vibrated at \ca~\SI{120}{\hertz}, for \SI{1}{h}, after a period of continuous settling in a rectangular tube of $L/b\approx 85$ and $\tilt=\SI{25}{\degree}$, recovering close to \SI{65}{\%} of the algae that could not be collected in the absence of vibration. This application did not just show the effectiveness of vibration, but also the potential to reduce the angle of settling elements. If particles are small enough to allow for buoyancy-viscous dominated flow, as the clarification rate in continuous operation and large $L/b$ is proportional to $\cos\tilt$, lower angles would further increase the efficiency of this type of equipment, whereas vibration might allow their operation below the angle of repose of fine particles with potential reductions of idle time for sediment removal.
If confinement is not so close or particle buoyancy is relatively large, then inertial effects on the flow may become dominant, and has been found to explain optimal angles in traditional \BES settlers for wastewater treatment between \SI{45}{\degree} and \SI{55}{\degree}, where higher angles tend to render the system least stable~\citep{Chang19PT}.
%
Relatively recent patented inventions related to this topic are the application of ultrasonic vibrations to the settling units~\citep{Sun12GP}, mechanical vibrations~\citep[][presents a modular system of inclined plates that can vibrate]{Kim13GP,Zhou13GP} and a system that activated an electrically-actuated vibration system on a tube array according to the effective thickness of the sediment layer.
\RB{The cost of vibration is, in principle, modest. Assuming a \SI{2}{kW} vibrating motor and considering one of them every 25 settling cells~\citep[e.g., as in the sketch presented in][]{Zhou12AMaM}, an unit energy cost of 75~USD/MWh yields an energy cost of the order of 10,000~USD/year for a 200-cell settler.}
In a mineral processing context, it is noted that dealing with separation of clays poses a strong restriction to the angle of the settler. \citet{Zahavi75IECPDaD} reported a minimum angle of \SI{43}{\degree} that allows clays to slide down Perspex planes, which are indeed very smooth surfaces~\citep[some complementary data can be found in][]{Al-Hashemi18PT}.

An alternative to the application of mechanical vibration is to apply it to the fluid rather than (or in addition to) the settler. Also for algae harvesting purposes, \citet{Hincapie15BP} used the principle of acoustophoresis, consisting of a net force on particles of certain size range, from antinodes to nodes of the mechanical excitation field (thus inducing agglomeration), in combination with a countercurrent \BES cell. In their experiments, they maximized algae flocs (primary size between \SI{3.7}{\micro m} and \SI{7.8}{\micro m}) with a resonant frequency near \ca \SI{1.74}{kHz} at an optimal angle of \SI{50}{\degree}. This resulted on peak filtration efficiencies up to about 7~times those without applying the ultrasound field.

A caveat to the principle of decreasing $\tilt$ and $b$ to increase $(L/b)\cos\tilt$ in buoyancy-viscous dominated flows when flocs rather than primary particles are to be separated is that the mean flow velocity (positive upwards in countercurrent settlers) may cause drag to preclude settled particles to report to the overflow or to induce particle accumulation at the bottom of the settling element~\citep{Driscoll08book}. This might pose a restriction to minimum clearance values, depending on the floc characteristics, including density and fractal dimension~\citep{Adelman13EES}. To this purpose, technology such as self-cleansing systems by moving robots within the settler~\citep{Nw21web} \Ri{may} solve, in a number of cases of practical interest, the problem associated to stopping the equipment for flushing or backwashing.




To successfully adapt such a scheme for secondary treatment of overflow water from thickeners in mineral processing plants, tight monitoring, and control of overflow particles and suspension physicochemistry would be required if low angle with or without vibration is to be used. While zeta potential values as close as possible to the isoelectric point would be desirable to obtain faster settling aggregates, they may also be prone to form sludge layers with higher yield stress \citep[there are several examples of such an effect in mineral fine processing slurries, such as][and references therein]{Zhou01CES,Reyes18ME,Contreras20ME}. This needs to be solved for successful incorporation of this unit process separate from thickening.


\subsection{Optimal inlet conditions}

One of the known potential problems of \BES clarifier units is the non-uniform or inadequate settler design at the feed~\citep{Huang08WSaT,Lee15EER}. Poor feed may be explained by a number of factors, including internal pressure loss differences throughout settling elements due to inhomogeneities in the velocity field at the feed points or the overflow. In this regard, collection through orifices made to the settling elements have proven effective to promote overflow re-distribution~\citep{Baruth04book}. Using CFD, \citet{He08JoEE}, showed that it is possible to improve the design of the inlet chamber to increase total suspended solid removal from \SI{67}{\%} to \SI{77}{\%}, while increasing its loading capacity by a factor close to~3. \citet{Morin09patent} filed a patent with an adjustable inlet, thus bringing flexibility to feed conditions in the interest of maximizing homogeneity in countercurrent lamella arrays. \citet{Feldthusen14GP} proposed an improvement at the collection zone using a plate sump supported from above with a design with several outlets, thus facilitating flow distribution. \citet{XuYin16GP} combined the used of inclined plates with collection tubes for efficient sludge outlet and a design of the collecting overflow with a multiport clamping plate. The critical importance of the inlet design was also established in the numerical study by~\citet{Nguyen19JoWaET} \citep[complemented by][]{Dao19JoWaET}, who found a \SI{10}{\%} difference in the suspended solid removal efficiency comparing two sideways rotated V-shaped plate array in a settling tank (one mimicking a `$\langle$' symbol and another a `$\rangle$' one). Considering an inlet from the left, the shape pointing to the left causing a rather diverging flow at the left proved better than the $\rangle$-shaped pattern.

Also, using CFD simulations, \citet{Okoth08CEaPPI} showed the impact of feed connection on the velocity distribution of a lamella settler (feed from one side compared with the installation of a distribution nozzle). This work was later complemented from a numerical perspective by~\citet{Salem11WR}, who compared a tubular feed without transition with one with a nozzle and a third one with the nozzle surrounded by a cylinder, thus minimizing the boundary layer separation near the nozzle (in the three cases, countercurrent flow configuration, with simulation parameters set for two turbulence models were considered). The need for proper orientation complements the conclusion of~\citet{Fuchs14WSaT} on the importance of maximizing energy dissipation at the inlet to eliminate turbulence as much as possible to achieve laminar flow in the settling elements. 
%


\RB{The increasing presence of clays in mineral ore slurry streams represents a challenge both to energy efficiency and to water footprint in the mining industry~\citep{Contreras20ME,Lopez-Espejo22M}.} In the context of \BES clarifier units fed with clay minerals, the settling velocity, which is of central relevance both to thickeners and clarifiers, depends on key factors such as the clay type, the flocculant characteristics~\citep{Addai-Mensah07PT,Wang14ME}, pH~\citep{Zbik08JoCaIS} and the solution salinity~\citep{Ji13CaSAPaEA}. The aforementioned mineral variability is thus naturally passed to the variability on the settling velocity and to the rheological characteristics of the underflow (two examples related to fine minerals are depicted in~\citet{Reyes19ME} and \citet{Contreras20ME} to account for the effect of seawater and blends of fine minerals, respectively). A case, made by~\citet{Castillo19M} with a blend of copper sulphide tailing samples with and without kaolinite, using various salts dissolved in the water, reveals that samples with similar chord lengths obtained using focused beam reflectance measurements ---a proxy of the floc size distribution--- can have different settling velocities depending on the presence of such clay and the salt content at similar chord length, thus indicating the differences induced by the resulting aggregate shape factor. The same work confirms that the supernatant turbidity is a strong function of the water characteristic (either seawater or distilled) and that pre-shearing plays a central role in turbidity removal after settling tests in upright containers. This reality, together with the central role of hydrodynamic conditions, pose a challenge on the design of feed connections in \BES equipment where, as in the case of thickener feedwell design, requires to be capable of handling significant variability of mineral combination at the feed and thereby: (1) may require different additive dosing strategies and (2) would benefit from having high flexibility on the selection of additive injection points~\citep[as recommended by][to improve thickener feedwell design]{Fawell19MPaEMR} depending on the minerals and water condition at the inlet and feed flow conditions.

\subsection{New developments in settling elements}

An alternative line of research is related to improvements related to the design of single settling elements. In the context of stormwater treatment using packs of inclined conduits, \citet[][and references therein]{Bhorkar05JoEE} propose both rotating square elements (\SI{45}{\degree}) and fitting insertions parallel to the main flow axis to enhance turbidity removal efficiency, reporting improvements between \SI{3}{\%} and \SI{4}{\%} \citep[a similar approach has been pointed out more recently, using other settling element cross-sections, in][with considerably higher efficiency improvement]{Bhorkar21JoIWWA}. A related effort to settling element design modification has been made, for cross-flow settling using lamella plates, by \citet{He09JoHE}, who propose to attach ribs to their surface, thus creating channelization with subsequent eddies in the zone between the ribs that, in the case of relatively fine particles, has been found to enhance settling. \citet{Kompala17GP} proposes settling element designs based on the Boycott effect including conical elements apex above or mounted inside hydrocyclones.

\begin{figure}[!h]
    \centering
    \begin{tabular}{cc}
        (a) & (b) \\
         \includegraphics[width=7.5cm,keepaspectratio]{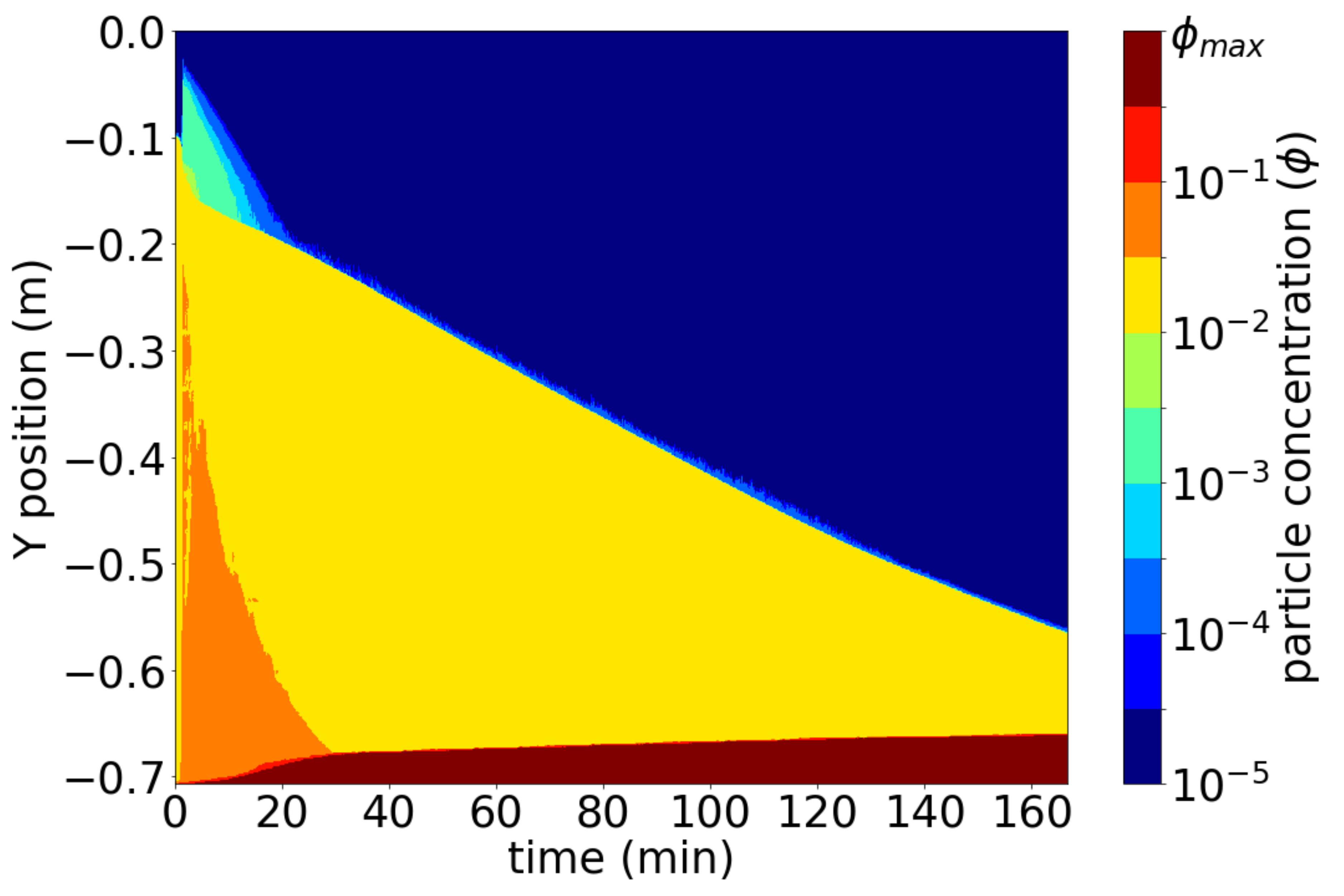}&
         \includegraphics[width=7.5cm,keepaspectratio]{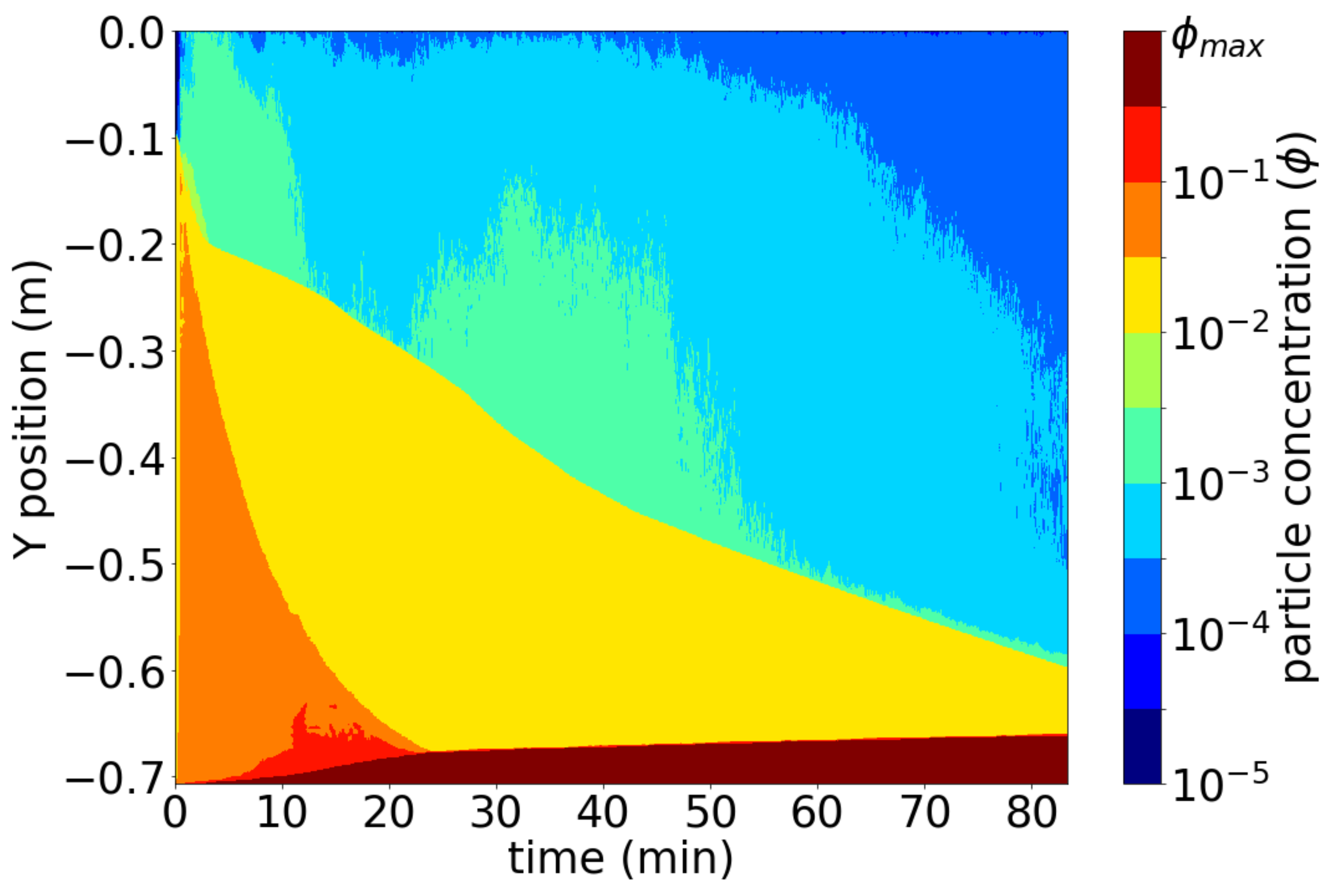}
    \end{tabular}
    \caption{Spatiotemporal diagram of particle volume concentration field for a fixed thin zone near the bottom plate, adapted from Figure~4 of~\citet{Reyes19M}. The settling cell has a tilting angle of \SI{45}{\degree}, a container height, and spacing of \SI{71}{cm} and \SI{5}{cm}, respectively. The suspension has a particle size of~\SI{10}{\micro m} in diameter and a solid phase density of \SI{2700}{kg/m^3}. Here, the so-called $Y$ position is the vertical coordinate, with $Y=0$ that corresponds to the top of the inclined cell. (a) No heating at the upper wall. (b) Imposed temperature step of \SI{30}{\celsius}, above the initial temperature of the suspension, at the upper wall. Note the difference between the time scales in both graphs.}
    \label{f:heated_spatiotemporal}
\end{figure}

A more recent development related to settling elements is the use of heat (very abundant, as solar energy, in many concentrator plants in desert locations) to enhance settling~\citep{Reyes19M}. Heat injection affects the structure of the clear fluid boundary layer \emph{via} natural convection in the downward-facing portion of settling elements. This sole mechanism, in the absence of particles, induces natural convection~\citep{Fujii72IJoHMT,Soong96IJoHaMT}. This process overlaps in a complex fashion with the particle-induced convection inherent to the Boycott effect. Figure~\ref{f:heated_spatiotemporal} shows a spatiotemporal diagram of two-dimensional two-phase model simulations of a batch experiment solved using the same finite volume scheme used in \citet{Reyes18ME}, where it is first noted that the sediment layer took roughly half of the time to form in the heated case, corresponding to an imposed temperature of \SI{30}{\celsius} above the initial temperature of the suspension at the upper wall over the suspension temperature (Figure~\ref{f:heated_spatiotemporal}b), than in the isothermal one. Here, the boundary between the clarified layer and the suspension (a traveling iso-concentration wave), denoted using false color by the yellow-blue and yellow-green interface in the absence and presence of heating, respectively, is significantly steeper when the upper wall is heated, as noted from the scale of the abscissa in both graphs.

Despite the encouraging results referred above, it is noted that the heated case is also related to some degree of particle contamination in the clarified layer (light-blue and green on top of yellow false-color compared in Figure~\ref{f:heated_spatiotemporal}b). \RB{Temperature is a critical aspect that may induce, if it is inhomogeneous within the settler, flow short circuit through the formation of gravity currents~\citep{Brown86thesis}. Even if the temperature feed is homogeneous, increasing it for a particular suspension may result on increased (and uncontrolled) feed flow that may cause a drop in the \TRE~\citep{Takata17WSaTWS}. It is therefore necessary to seek either for homogeneous temperature conditions inside the settler and to have proper knowledge of daily/seasonal variations of water temperature, so the performance can be adjusted accordingly or, if it cannot be controlled, be known beforehand.} Therefore, this approach, which is presently under development, requires very tight control both of feed conditions and heating. 

\section{Final remarks}



Water footprint minimization in mineral processing plants is among the top priorities of the mining industry. With that comes the need to fulfill the process objectives of extractive metallurgy. The present document has presented the use of \BES technology as a means to make better control of water recovery from thickeners by adding an independent unit process to control the quality of the overflow water (in the sense of suspended solid content). Figure~\ref{f:lamella-countercurrent} shows a lamella clarifier fitted with a feed tank where turbulence can be imposed and controlled. This gives an additional opportunity to design thickener overflow clarification strategies relying on reduced use of reagents, provided a thickener operation prioritizing settling velocity may be done at with lower flocculant consumption but at the expense of tighter control of dilution at the thickener. The details of such a tandem strategy are beyond the scope of the present paper, and certainly, a process engineering approach is required to explore the best simultaneous operational points. From this perspective, this problem resembles that of geometallurgy~\citep[reviewed by][]{Lishchuk20ME}, as it is inherently bonded to mineral and water characteristics.

The present paper shows a convergence of the parallel alleys that have marked the research in the last century which started with the discovery of~\citet{Boycott20N}, and the developments in the last 50~years concerning \BES equipment, mostly applied to wastewater. The PNK theory (based on mass balance concepts) matches the two-dimensional approach to set the minimum (infinitely wide) inclined plate length to host all particles of a specified size in a dilute suspension. This is to be tested in different cross-sectional geometries. This would help, in particular, to provide experimental support to what is suggested by~\citet{Ihle21PT} on the relevance of the cross-sectional shape on the footprint of~\BES equipment. It would therefore be interesting to combine both approaches to update already existing engineering models for system design.

\RB{Although the development of \BES technology is mature, and that there are a number of well established industrial producers of this kind of equipment, there is room for improvement of the overall clarification process, especially in light of increasing and variable fine and ultrafine contents in mineral concentrator plants. The interplay between the use of additives, the kind and size of particles to be separated, the concentration of particles and the dimensional characteristics of the settlers imply a diversity of TRE results. From the perspective of equipment design, several} improvement opportunities are yet to be explored in modern systems, including fine-tuning of settling element arrays (angle, spacing, and the potential inclusion of mechanical attachments for controlled vibration), to optimize inlet conditions to make the most of the settling element array and to optimize the design of settling elements. At this point, there is also an opportunity to use solar energy, which is particularly abundant in several mining operations of Chile, Australia, and elsewhere, to further reduce the footprint. Nonetheless, in this case, to alter design means to re-visit the physics behind the Boycott effect, which represents a challenge on fluid dynamics and the physics of suspensions. It is also noted that, despite the present update of the settling technology equipment, general observations, including the interlink between suspension characteristics and optimal settler design, still hold and justify the existence of \BES settler design from a research and development perspective.

Although the use of experimental techniques for system modeling is irreplaceable, the opportunities from existing modern computational capacity result in significantly reduced system development times. So far, this has particularly benefited the design of settling basins with attached arrays on inclined settling elements using low-computational-cost turbulence closures such as $k$-$\epsilon$ and $k$-$\omega$, which are included in most commercial CFD packages. Comparatively, little work has been done concerning the use of high-resolution simulations to analyze the details of the flow, where such turbulence closures are less useful. This can be particularly challenging, given the potential width of the clear fluid boundary layer (in the case of subcritical settlers) configuring laminar flow, the potential particle size. It can have broad variations in the same sample and potentially multiple flow regimes: commonly turbulent flow near the inlet of the settler, transitional flow near the entry of the settling elements, and laminar flow within them. The opportunities that offer carefully chosen and configured Eulerian-Eulerian schemes (\ie, both solid and liquid phases seen as continuous) or Eulerian-Lagrangian setups allow not just to anticipate hydrodynamic instabilities that may cause detriment on performance (and to have a detailed knowledge of the system SOR under the strongly variable plant feed and reagent conditions described above) but also to reasonably model the downstream sludge flow.









\section*{Acknowledgements}

The authors gratefully acknowledge support from the Department of Mining Engineering of University of Chile and the Chilean National Agency for Research and Development through PIA Grant AFB180004, Fondecyt Project 1211044, and National Doctoral Scholarship 21190656. CI also acknowledges support from the Vice-rectory of Research and Development Univ. Chile, Project code ENL11/20. CA acknowledges support of Nordita and the Swedish Research Council Grant No. 2018-04290; Nordita is partially supported by Nordforsk.

\appendix

\section{Scale of clear fluid layer in subcritical mode of operation (negligible inertia)}\label{a:scaling-derivation}

Consider the scales $\uscale$, $\vscale$, $\longscale$, and $\blscale$. Here, $\uscale$ and $\vscale$ are velocity scales parallel and perpendicular to the upper wall (facing downwards) at a (small) distance $\blscale$ from the upper wall. The latter distance is assumed much thinner than the longitudinal length scale $\longscale$ so, to first order, $\blscale\ll\longscale$, as confirmed by numerous experiments~\citep[][and references therein]{Hill77IJoMF}. Continuity near the boundary layer implies that 
$
    \frac{\uscale}{\longscale}\sim \frac{\vscale}{\blscale}.
$
If the tube is tilted from the vertical, then $\longscale\sim\ho$, where $\ho$ is the projected height of the inclined tube.

Assuming that the flow at the boundary layer is laminar, then the force density  $F_{d,f}$ balances the viscous force per unit volume, and therefore
\begin{equation}
    \frac{\muf\uscale}{\blscale^2}\sim g\Cv_0(\rhos-\rhof).
\end{equation}
Further, assuming that $\vscale$ is on the order of the settling velocity, $\vscale\sim \WI$, implies that the longitudinal velocity and the thickness of the boundary layer verify~\citep[see also][]{Blanchette03thesis}:
\begin{subequations}\label{e:u-delta-scales}
\begin{align}
    \frac{\uscale}{\WI} &\sim \rGR^{1/3}\\
    \frac{\blscale}{\ho}&\sim \rGR^{-1/3},\label{e:blscale}
\end{align}
\end{subequations}
Using $\WI=\wSt$, the dimensionless parameter $\rGR$ corresponds to~\eqref{e:rGR}. Also, under this definition of $\WI$, $\rGR=18(\ho/d)^2\Cv_0$. From this boundary layer perspective, noting that $\ho\sim x$, where $x$ is a longitudinal coordinate parallel to the upper wall), then, from~\eqref{e:blscale}, $\blscale\propto x^{1/3}$.

\section*{List of symbols}
\nopagebreak
\begin{longtable}[l]{@{}ll}
$\delta$& distance between control volume and inner diameter of lower portion of settler\\
$\Cv$   & volume fraction (dimensionless)\\
$\Cv_0$ & feed volume fraction (dimensionless)\\
$\Cv_U$ & underflow volume fraction (dimensionless)\\
$\rGR$  & Grashof-Reynolds number ratio (dimensionless)\\
$\muf$  & fluid phase dynamic viscosity (mass\,length$^{-1}$\,time$^{-1}$)\\
$\EG$   & settling cell efficiency (dimensionless)\\
$\tilt$ & inclination angle, measured from the horizontal\\
$\partial\Omega$& boundary of the control volume (length$^2$)\\
$\Omega$& control volume (length$^3$)\\
$\wparam$ & $\vset/\vmean$ (dimensionless)\\
$\rhof$    & fluid phase density (mass/length$^3$)\\
$\tilde{\rho}_s$ & solid mass per unit volume (mass/length$^3$)\\
$\rhos$    & solid phase density (mass/length$^3$)\\
$A_E$   & projected area of settler (length$^2$)\\
$b$     & height of settling element, measured from the base (length)\\
$D$     & inside diameter of pipe (length)\\
$d$     & hexagon side in Figure~\ref{f:hex-pack} (length)\\
$d_p$   & particle diameter (length)\\
$\bv{\hat{F}}_{d,p}$ & drag force on a sediment particle (mass length/time$^2$) \\
$\bv{F}_{d,f}$ & force density by the sediment on the fluid (mass length$^{-2}$ time$^{-2}$) \\
$\mathit{Gr}$ & Grashof number (dimensionless)\\
$\bv{g}$& acceleration of gravity vector (length/time$^2$)\\
$g$     & magnitude of acceleration of gravity vector (length/time$^2$)\\
$h$     & height of suspension (length)\\
$\nhat$ & unit normal pointing outward the control volume (dimensionless)\\
$\longscale$ & longitudinal length scale (length)\\
$L$     & settling element (conduit) length (length)\\
$n_p$ & number of sediment particles per unit volume (length$^{-3}$) \\
SOR     & surface overflow rate (length/time)\\
$\qmax$ & maximum volume flow per unit width to avoid particles at the overflow (length$^2$/time)\\
$\qPNK$ & volume flow per unit width predicted by PNK theory (length$^2$/time)\\
$\NRe$ & Reynolds number (dimensionless)\\
$\ReCell$ & stability control parameter (dimensionless)\\
$\ReCellb$ & critical stability parameter (dimensionless)\\
$\dimClar$& clarification rate per unit depth (length$^2$/time)\\
$s$     & clarification rate per unit area (length/time)\\
$\SY$    & dimensionless clarification parameter\\
$\turb_0$ & turbidity at the feed (turbidity units)\\
$\turb_{\OF}$ & turbidity at the overflow (turbidity units)\\
$\TRE$  & turbidity removal efficiency (dimensionless)\\
$t$     & time (time)\\
$\vmean$& mean flow velocity magnitude (length/time)\\
$\bv{u}_s$ & solid phase velocity vector (length/time)\\
$u_{y,\text{cp}}$ & velocity at the plane of symmetry\\
$\uscale$ & velocity scale parallel to the slope (length/time)\\
$\vscale$ & velocity scale normal to the slope (length/time)\\
\Vp  & particle volume (length$^3$)\\
$\bv{\vset}$ & settling velocity vector (length/time)\\
$\vset$ & settling velocity vector magnitude (length/time)\\
$W$     & domain width (length)\\
$\wSt$  & Stokes settling velocity (length/time)\\
$x$     & horizontal coordinate (Figure~\ref{f:PNK_schematic}) or longitudinal coordinate (length)\\
$y$     & coordinate normal to settling element main axis (length)\\
$z$     & vertical coordinate\\
\end{longtable}



\end{document}